\documentstyle[12pt,epsf]{article}

\newlength{\dinwidth}
\newlength{\dinmargin}
\newcommand{\resection}[1]{\setcounter{equation}{0}\section{#1}}

\newcommand{\appsection}{\addtocounter{section}{1} \setcounter{equation}{0}
                         \section*{Appendix}}
\renewcommand{\theequation}{\thesection.\arabic{equation}}
\begin{document}
\vspace*{4cm}
\begin{center}
  \begin{Large}
  \begin{bf}
NEW VECTOR BOSONS  IN THE ELECTROWEAK SECTOR: A RENORMALIZABLE MODEL
WITH DECOUPLING\\
  \end{bf}
  \end{Large}
  \vspace{5mm}
  \begin{large}
R. Casalbuoni,  S. De Curtis and D. Dominici\\
  \end{large}
Dipartimento di Fisica, Universit\`a di Firenze\\ I.N.F.N., Sezione di
Firenze\\
  \vspace{5mm}
  \begin{large}
M. Grazzini\\
  \end{large}
Dipartimento di Fisica, Universit\`a di Parma \\ I.N.F.N., Gruppo
collegato di Parma
  \vspace{5mm}
 \vspace{5mm}
\end{center}
  \vspace{2cm}
\begin{center}
University of Florence - DFF 272/02/97
\end{center}
\vspace{1cm}
\noindent
%$^*$ Partially supported by the Swiss National Foundation
\def\spin{{2}}
\def\uu{\dd{u^2}}
\def\vv{\dd{v^2}}
\def\lq{\left [}
\def\rq{\right ]}
\def\qq{Q^2}
\def\ct{c_{\theta}}
\def\ctz{c_{{\theta}_0}}
\def\st{s_{\theta}}
\def\stz{s_{{\theta}_0}}
\def\cpsi{\cos\psi}
\def\spsi{\sin\psi}
\def\sf{s_{\varphi}}
\def\cf{c_{\varphi}}
\def\dmus{\partial^{\mu}}
\def\dmu{\partial_{\mu}}
\def\LL{{\cal L}}
\def\BB{{\cal B}}
\def\Tr{{\rm Tr}}
\def\gp{g'}
\def\gs{g''}
\def\ggs{\frac{g}{\gs}}
\def\eps{{\epsilon}}
\def\f{\frac}
\def\L{{W^{\prime }}_L}
\def\R{{W^{\prime }}_R}
\newcommand{\be}{\begin{equation}}
\newcommand{\ee}{\end{equation}}
\newcommand{\bea}{\begin{eqnarray}}
\newcommand{\eea}{\end{eqnarray}}
\newcommand{\nn}{\nonumber}
\newcommand{\dd}{\displaystyle}
\newpage
\thispagestyle{empty}
\begin{quotation}
\vspace*{5cm}

\begin{center}
\begin{bf}
  ABSTRACT
  \end{bf}
\end{center}
\vspace{1cm}
\noindent
A linear realization of a model of dynamical electroweak symmetry
breaking  describing additional heavy vector bosons is proposed. The
model is a $SU(2)_L\otimes U(1)\otimes SU(2)_{L}^\prime
\otimes SU(2)_{R}^\prime$ gauge theory, breaking at some high scale
$u$ to $SU(2)_{weak}\otimes U(1)_Y$ and  breaking again in the
standard way at the electroweak scale $v$ to $U(1)_{em}$. The model is
renormalizable and reproduces the Standard Model in the limit
$u\to\infty$. This decoupling property is shown to hold also at the
level of radiative corrections by computing, in particular, the
$\epsilon$ parameters.

  \vspace{5mm}
\noindent
\end{quotation}
\newpage
\setcounter{page}{1}

\resection{Introduction}

Existing experimental data confirm with great accuracy the Standard
Model (SM) of the electroweak interactions. Therefore, only extensions
which smoothly modify the SM predictions are still conceivable. The
Minimal Supersymmetric Standard Model (MSSM) \cite{mssm} is the most
favorite one because, in addition to many other interesting features,
in the heavy limit (that is the limit in which all superpartner masses
become heavy) decoupling holds and the MSSM becomes at low energy
indistinguishable from the SM with a low Higgs mass \cite{barbieri}.
However, this decoupling property is not peculiar of the MSSM; for
instance, it is satisfied also in the non supersymmetric two Higgs
model \cite{haber}.
 There are also examples of
dynamical symmetry breaking  schemes satisfying such a property. In
fact, in a previous paper \cite{degene} we have considered a model
(degenerate BESS) of a strong electroweak breaking sector  describing,
besides the  standard $W^\pm$, $Z$ and $\gamma$ vector bosons, two new
triplets of spin 1 particles, $V_ L$ and $V_R$. These new states are
degenerate in mass if one neglects their mixing to the ordinary vector
bosons. The description of the model was based on a non-linear gauged
$\sigma$-model and we refer to \cite{degene} for more details. The
interest in this scheme was due to the fact that it decouples: in the
limit of infinite mass of the heavy vector bosons  one gets back the
Higgsless SM. This is a rather non trivial property because one is
dealing with a non-linear theory with couplings increasing with the
heavy masses. In fact, the decoupling originates from an accidental
global symmetry that the model possesses when the gauge couplings are
turned off. This is also the symmetry from which the quasi-degeneracy
of the heavy vector states arises.

The original philosophy of the non-linear version was based on the
idea that the non-linear realization would be the low-energy
description of some underlying dynamics giving rise to the breaking of
the electroweak symmetry. In  a recent paper \cite{linear} we have
suggested a linear realization of this model,  which might appear as
based on a completely different standpoint. We are thinking of a
scenario very close to the one arising in technicolor \cite{tc} and in
generalizations as non-commuting technicolor models \cite{nctc}, where
one has an underlying strong dynamics producing heavy Higgs composite
particles. In this sense we are trying to describe the theory at the
level of its composite states, vectors (the new heavy bosons), and
scalars (Higgs bosons). That is, we are looking at a scale in which
the Higgs bosons are yet relevant degrees of freedom. The advantage is
to deal with a renormalizable theory. By that, one is able to discuss
the decoupling at the level of radiative corrections.

The model is a $SU(2)_L\otimes U(1)\otimes SU(2)_{L}^\prime \otimes
SU(2)_{R}^\prime$ gauge theory, breaking at  some high scale $u$ to
$SU(2)_{weak}\otimes U(1)_Y$ and  breaking again in the standard way
at the electroweak scale $v$ to $U(1)_{em}$. In this paper we will
show that this  model in the limit of  large $u$ decouples also at one
loop level, and consequently that the high-energy physics  is not
relevant at the LEPI scale. Therefore the model we present is identical
to the SM in its low energy manifestations, although at higher
energies the differences can be rather dramatic \cite{degene, lhc}.

To show the decoupling we have concentrated on the observables which
are relevant to LEPI physics, that is, on the so called $\epsilon$
parameters \cite{altarelli} (or the corresponding $S$, $T$, $U$
parameters \cite{peskin}). The universality property holds in the
model, so we need only to consider the parameters $\epsilon_i$,
$i=1,2,3$. We have performed the calculation of the corrections coming
from the heavy sector of the model, following the scheme outlined in
\cite{barbieri}. In this scheme one has to evaluate the self-energy
corrections to the standard gauge boson propagators, the vertex
corrections to $Z\to e^+ e^-$ and the corrections to the Fermi
coupling constant. At the end one collects all the various
contributions together in order to reconstruct the physical
quantities. For this reason we have not studied in detail the
renormalization property of the model, but rather we have evaluated
the different corrections in dimensional regularization and in the
unitary gauge. In fact, as it must be, all the ultraviolet divergences
cancel out when we evaluate the $\epsilon_i$ parameters. To make the
calculations easier we have performed a particular transformation on
the gauge parameters, in such a way to make the SM limit transparent.
Also, we have chosen to work with mass eigenstates, because this makes
those couplings, which increase with the heavy mass, to appear in only
two sectors of the model. The first one is the Higgs sector, which,
however, is not relevant in our calculations barring the standard
hierarchy problem. The second one is the heavy-Higgs heavy-vector
sector. This is shown to be harmless in the text. By following the
previous procedure we show explicitly that no contribution to the
$\epsilon_i$ parameters survive in the heavy mass limit, proving the
decoupling of the model. This property appears to be strictly related
to the absence of couplings increasing with the heavy scale in the
light-light sector and in the heavy-light one (except for the Higgs
case, as mentioned before). In fact, this is what one would expect
from the Appelquist-Carazzone theorem \cite{AC}. However, the absence
of these couplings is evident only in the unitary gauge, where the
cancellations among the different contributions to the observable
quantities are far from being trivial.

In Section 2 we will review  the linearized version of the model. In
Section 3 the scalar potential and the symmetry breaking are  studied.
In Section 4 the spectrum of gauge vector bosons  and their
interactions with the fermions are analyzed, showing in particular how
the SM relations are obtained in the $u\to\infty$   limit. In Section
5 we perform the calculation at  tree level of the $\epsilon $
parameters in the $u\to\infty$ limit, showing that they are  of
$O(v^2/u^2)$. General formulas for the $\epsilon $ parameters in terms
of vacuum polarization amplitudes for the $W$, $Z$ and $\gamma$,
contributions to vector and axial-vector form factors at the $Z$ pole
in the $Z e^+e^-$ vertex and one loop corrections to the $\mu$ decay
amplitude are given in Section 6. Explicit one loop results for the
vacuum polarization amplitudes in the $u\to\infty$ limit are given in
Section 7. In Section 8 we derive some one loop general result for
box,  vertex  and  fermion self-energy amplitudes. In Section  9 one
loop corrections to $G_F$ in the $u\to\infty$ limit are considered,
showing that they vanish. In Section 10 we show that one loop
corrections to vector and axial-vector form factors at the $Z$ pole
vanish in the same limit. In Appendix we give the explicit expressions
of the relevant Higgs and gauge boson interaction terms.

\resection{The Model}

The model \cite{linear}, that  we briefly recall  here, is based on a
gauge group $SU(2)_L\otimes U(1)\otimes SU(2)_{L}^\prime
\otimes SU(2)_{R}^\prime$ and has a scalar sector consisting of scalar
fields belonging to the following representations of the group
$SU(2)_L\otimes SU(2)_R\otimes SU(2)_{L}^\prime \otimes
SU(2)_{R}^\prime$
\be
\tilde L\in (\spin,0,\spin,0),~~~~~
\tilde U\in (\spin,\spin,0,0),~~~~~
\tilde R\in (0,\spin,0,\spin),
\ee
that is with transformation properties
\be
{\tilde L}^\prime = g_L\tilde  L h_L,~~~~~ {\tilde U}^\prime =
g_L\tilde U g_R^\dagger, ~~~~ {\tilde R}^\prime = g_R\tilde R h_R,
\ee
where
\bea
& & g_L\in {SU(2)_L},~~~~~ g_R\in {SU(2)_R},\nn\\
 & &h_L\in SU(2)_L^\prime,~~~~~
h_R\in SU(2)_R^\prime.
\eea
We will see that with this system of scalar fields it is possible to break the gauge
symmetries through the following chain
\be
\matrix{
SU(2)_L\otimes U(1)\otimes SU(2)_{L}^\prime \otimes SU(2)_{R}^\prime\cr
\downarrow u\cr
SU(2)_{\rm weak}\otimes U(1)_Y\cr
\downarrow v\cr
U(1)_{\rm em}}\label{1.4}
\ee
The two breakings are induced by the expectation values $\langle\tilde L\rangle=
\langle\tilde R\rangle=u$ and $\langle\tilde U\rangle=v$ respectively.
The first two expectation values make the breaking $SU(2)_L\otimes
SU(2)'_L\to SU(2)_{\rm weak}$ and $U(1)\otimes SU(2)'_R\to U(1)_Y$,
whereas the second breaks in the standard way $SU(2)_{\rm weak}\otimes
U(1)_Y\to U(1)_{\rm em}$. In the following we will assume that the
first breaking corresponds to a scale $u\gg v$.

Proceeding in a completely standard way, we can build up
covariant derivatives with respect to the
local
$SU(2)_L\otimes U(1)\otimes SU(2)_{L}^\prime \otimes SU(2)_{R}^\prime$
\bea
& &D \tilde L =\partial \tilde L +
i g_0 \frac{{\vec \tau}}{2}\cdot{\vec W} \tilde L
-i g_2 \tilde L\frac{{\vec \tau}}{2}\cdot{{\vec V}_L},  \nn\\
& &D \tilde R =\partial\tilde R +i  g_1 \frac{{ \tau_3}}{2}{Y}\tilde R
-i g_3\tilde R\frac{{\vec \tau}}{2}\cdot{{\vec V}_R},  \nn\\
& &D\tilde U =\partial\tilde U +
i g_0 \frac{{\vec \tau}}{2}\cdot{\vec W} \tilde U
- i  g_1\tilde U\frac{{ \tau_3}}{2}{Y},
\label{covariant}
\eea
where ${\vec V}_L~({\vec V}_R)$
are the  gauge fields in $SU(2)_L^\prime ~
(SU(2)_R^\prime)$,
 with the corresponding gauge couplings  $g_2$, and
$g_3$, whereas $g_0$, $g_1$, are the gauge couplings of the $SU(2)_L$ and $U(1)$
gauge groups respectively.

This model contains, besides the standard Higgs sector given
by the field $\tilde U$, the additional scalar fields $\tilde L$ and $\tilde R$.

The Lagrangian for the kinetic terms of these scalar fields
is given by
\be
{\cal L}^h = \frac 1 4 \lq Tr (D_\mu \tilde U)^\dagger  (D^\mu\tilde
U) +Tr (D_\mu \tilde L)^\dagger  (D^\mu \tilde L) +Tr (D_\mu \tilde
R)^\dagger  (D^\mu \tilde R) \rq.
\ee
We have then to discuss the scalar potential which is supposed to
break  the original symmetry down to the $U(1)_{\rm em}$ group.
The most general potential invariant with respect to the group
 $SU(2)_L\otimes SU(2)_R\otimes SU(2)_{L}^\prime \otimes SU(2)_{R}^\prime$
 is given by
\bea
&V(\tilde U,\tilde L,\tilde R)=&\mu_1^2 Tr ({\tilde L}^\dagger \tilde
L) +\frac { \lambda_1} {4}[ Tr ({\tilde L}^\dagger \tilde L)]^2
+\mu_2^2 Tr ({\tilde R}^\dagger \tilde R) +\frac {\lambda_2} {4} [Tr
({\tilde R}^\dagger \tilde R)]^2\nn\\ & &+m^2  Tr ({\tilde U}^\dagger
\tilde U) +\frac h 4 [Tr ({\tilde U}^\dagger \tilde U)]^2 +\frac {f_3}
{2} Tr ({\tilde L^\dagger}\tilde L) Tr ({\tilde R^\dagger}\tilde
R)\nn\\ &&+\frac {f_1}{2} Tr ({\tilde L^\dagger}\tilde L ) Tr ({\tilde
U^\dagger}\tilde  U) +\frac {f_2}{2} Tr ({\tilde R^\dagger}\tilde R)
Tr ({\tilde U^\dagger}\tilde U).
\eea

In the following we will also require, for the scalar potential,
the discrete symmetry $L\leftrightarrow R$, which
implies
\bea
&&g_3=g_2,\nn\\ &&\mu_1=\mu_2=\mu,\nn\\
&&\lambda_1=\lambda_2=\lambda,\nn\\ &&f_1=f_2=f.
\eea

The total Lagrangian is obtained by adding the kinetic terms for
the gauge fields:
\be
{\cal L}={\cal L}^h-V(\tilde U,\tilde L,\tilde R) +{\cal
L}^{kin}(W,Y,V_L,V_R),
\ee
where
\bea
{\cal L}^{kin}(W,Y,V_L,V_R)&=
&\frac 1 2 {\rm tr}[F_{\mu\nu}(W)F^{\mu\nu}(W)]+
\frac 1 2 {\rm tr}[F_{\mu\nu}(Y)F^{\mu\nu}(Y)]\nn\\
&+&\frac 1 2 {\rm tr}[F_{\mu\nu}(V_L)F^{\mu\nu}(V_L)]+
\frac 1 2 {\rm tr}[F_{\mu\nu}(V_R)F^{\mu\nu}(V_R)].\label{lkin}
\eea
Notice that, when neglecting the gauge interactions,  the Lagrangian
is invariant under an  extended symmetry corresponding to
$(SU(2)_L\otimes SU(2)_R)^3$. In fact, in this case, we are free to
change any of the fields $\tilde U$, $\tilde L$, $\tilde R$ by an
independent transformation of a group $SU(2)_L\otimes SU(2)_R$
\cite{degene}. As far as the fermions are concerned they transform as
in the SM with respect to the group $SU(2)_L\otimes U(1)$,
correspondingly the Yukawa terms are built up exactly as in the SM.

\resection{The scalar potential}

Let us parameterize the scalar fields as
\be
\tilde L=\rho_L  L,~~~~~ \tilde R= \rho_R  R, ~~~~~
\tilde U=\rho_U  U,
\ee
with    ${L}^\dagger {L}=I$,   ${R}^\dagger{R}=I$
 and ${U}^\dagger  {U}=I$.

The scalar potential after these transformations can be rewritten
as
\bea
V(\rho_U,\rho_L,\rho_R)&=& 2 \mu^2 (\rho_L^2+\rho_R^2)+
\lambda (\rho_L^4+\rho_R^4) +2 m^2 \rho_U^2 +
 h \rho_U^4\nn\\
&+& 2 f_3 \rho_L^2\rho_R^2 + 2 f \rho_U^2(\rho_L^2+\rho_R^2).
\eea

To study the minimum conditions, let us consider the
first derivatives of the potential
\be
\frac {\dd \partial V}{\dd\partial\rho_L}=
4 \rho_L (\mu^2 +\lambda \rho_L^2 + f_3 \rho_R^2 + f \rho_U^2),
\ee
\be
\frac {\dd \partial V}{\dd\partial\rho_R}=
4 \rho_R (\mu^2 + \lambda \rho_R^2+ f_3 \rho_L^2+ f \rho_U^2),
\ee
\be
\frac {\dd \partial V}{\dd\partial\rho_U}=
4 \rho_U (m^2 + h \rho_U^2 + f (\rho_L^2 + \rho_R^2)).
\ee

By substituting the vacuum expectation values $<\rho_U>=v$ and
$<\rho_L>=<\rho_R>=u$, the minimum conditions  are
\be
\mu^2+ (f_3+\lambda)u^2 + f v^2=0,
\label{2.6}
\ee
\be
m^2+ 2fu^2 +h v^2=0.
\label{2.7}
\ee

From the second derivatives of the potential we get the mass matrix
for the three Higgs particles
\be
8 \pmatrix{\lambda u^2&f_3 \uu&f u v\cr f_3\uu&\lambda \uu&f uv\cr
fuv&fuv&h\vv}.
\ee
The mass eigenvalues are
\bea
M^2_{\rho_U}&=&4\Big[(f_3+\lambda)\uu +h \vv -\sqrt{8\uu\vv f^2+
((f_3+\lambda)\uu- h\vv)^2}\Big],\nn\\ M^2_{\rho_L}&=&8\lambda \uu (1-
\frac {f_3}{\lambda}),\nn\\ M^2_{\rho_R}&=&4\Big[(f_3+\lambda)\uu +h
\vv +\sqrt{8\uu\vv f^2+ ((f_3+\lambda)\uu- h\vv)^2}\Big].
\eea
Let us comment on the limitations on the parameters coming from the
study of the positivity of the eigenvalues. Adding the requirement
$u^2>0$, $v^2>0$, with the hypothesis $m^2,\mu^2<0$ together with
$\lambda,h>0$ for the boundedness of the potential, we finally get
\be
\lambda -f_3>0,~~~~h>f \frac {m^2}{\mu^2},
\ee
and
\be
\lambda +f_3> 2 f \frac {\mu^2}{m^2}~~for~f>0,~or
\ee
\be
\lambda +f_3> 2 \frac {f^2} {h} ~~for~f<0.
\ee

As shown in  \cite{linear} the limit $u\to\infty$ gives the SM with a
Higgs field light with respect to the scale $u$, with the following
redefinition of the gauge coupling constants
\bea
\frac 1 {g^2}&=&\frac 1 {g_0^2}+\frac 1 {g_2^2},\nn\\
\frac 1 {g'^2}&=&\frac 1 {g_1^2}+\frac 1 {g_2^2}.\label{2.10}
\eea
At the lowest order in the large $u$ expansion we get for the Higgs
mass eigenvalues
\bea
M^2_{\rho_U}&\sim& 8 \vv (h-2\f {f^2}{f_3+\lambda}),\nn\\
M^2_{\rho_L}&\sim&8 \uu (\lambda-  {f_3}),\nn\\ M^2_{\rho_R}&\sim&
8u^2(\lambda  +  {f_3}).
\eea

The scalar potential, after the shift $\rho_L\to \rho_L+u$,
 $\rho_R\to \rho_R+u$ and  $\rho_U\to \rho_U+v$, and by substituting
$m^2,~\mu^2$ as functions of the other parameters by using the minimum conditions
 (\ref{2.6}) and (\ref{2.7}),
becomes
\bea
V(\rho_U,\rho_L,\rho_R)&=& 4 h v^2 \rho_U^2+ 8 f u v \rho_U (\rho_L+\rho_R)
+ 4 \lambda u^2 (\rho_L^2+\rho_R^2)+8 f_3 u^2 \rho_L\rho_R \nn\\
&+& 4 h v \rho_U^3+ 4 \lambda u (\rho_L^3+\rho_R^3)+
4 f u \rho_U^2 (\rho_L+\rho_R)+4 f v \rho_U(\rho_L^2+\rho_R^2)\nn\\ &+&
4 f_3 u (\rho_R\rho_L^2+\rho_L\rho_R^2)+h \rho_U^4 +\lambda (\rho_L^4+\rho_R^4)
\nn\\
&+& 2 f_3 \rho_L^2\rho_R^2 + 2 f \rho_U^2(\rho_L^2+\rho_R^2).
\label{pot}\eea

Since we will not be interested in the Higgs self-interactions, we
will not give the explicit expression of the scalar potential in terms
of the mass eigenstates. It can be easily obtained by  using the
matrix ${\bf H}$ which transforms the fields $(\rho_L,\rho_R,\rho_U)$
appearing in eq. (\ref{pot}) into the Higgs eigenstates which we will
keep on calling in the same way. At the first order in
\be
r=\f {v^2}{u^2}\f{g^2}{g_2^2},
\label{r-parameter}
\ee
 we get
\be
{\bf H}^{-1}=
\pmatrix{
\dd{\f 1{\sqrt{2}}} & \dd{\f 1{\sqrt{2}}}(1-\dd{\f{q^2}{\sf^2}}r)
 & -\dd{\f q {\sf}}\sqrt{r}\cr
 -\dd{\f 1{\sqrt{2}}} & \dd{\f 1{\sqrt{2}}}(1-\dd{\f{q^2}{\sf^2}}r)
 & -\dd{\f q {\sf}}\sqrt{r}\cr
 0 & \dd{\f q {\sf}}\sqrt{2 r} & 1-\dd{\f{q^2}{\sf^2}}r  \cr
},
\ee
with
\be
q=\f f {f_3+\lambda},
\ee
and
\be
s_\varphi=\f g {g_2},
\ee
 in terms of which
\be
g_0=\frac g {\cf},
\ee
and
\be
\frac {g_1} {g_0}=\frac {\cf \st}{\sqrt{P}},
\ee
where
\be
\tan\theta=\frac {g'} g,
\ee
and
\be
P=\ct^2-\sf^2 \st^2.
\ee

\resection{Gauge vector boson spectrum and interactions}

The  vector boson mass spectrum can be studied in the unitary
gauge $ U=  L=  R =I$ by shifting
the scalar fields as
$\rho_U\to \rho_U +v$, $\rho_{L,R}\to \rho_{L,R} +u$.
We get
\bea
\LL^h&=&\frac 1 2 \left   [ (\dmu \rho_L)^2 + (\dmu \rho_R)^2  +
(\dmu \rho_U)^2 \right ]\nn\\
&+&\frac 1 8  \{ (\rho_L+u)^2  [ g_0^2 (W_3^2 + 2 W^+W^-)
- 2 g_0 g_2 (W_3V_{3L}+ W^-V_{L}^{ +}
+W^+V_{L}^{ -})\nn\\ &+&g_2^2 (V_{3L}^{ 2} + 2 V_L^{ +}V_L^{ -})
]\nn\\ &+&(\rho_R+u)^2  [g_1^2 Y^2 -2 g_1 g_2 V_{3R}Y+ g_2^2 (V_{3R}^{
2} +2 V_R^{+} V_R^{ -})] \nn\\ &+& (\rho_U+v)^2 [g_0^2 (W_3^2 + 2
W^+W^-)-2 g_0 g_1 W_3 Y + g_1^2 Y^2 ] \}.
\label{3.1}
\eea

Here we are  interested in the mass matrices for large mass
eigenvalues of ${V}_{L,R}$, $\rho_{L,R}$. First of all, it turns out
to be convenient to re-express the results in terms of the parameters
$g$ and $g'$ defined in eq. (\ref{2.10}). In fact, as we have said,
these are the relevant parameters in the limit $u\to\infty$. Let us
study the mass eigenvalues of the vector bosons in the charged and in
the neutral sector.

\hfill\break\medskip\noindent
\underbar{Charged gauge sector}
\hfill\break\medskip\noindent
The fields ${V}^\pm_R$ are unmixed and their mass is given by
\be
M^2_{V_R}=\frac 1 4 g_2^2 u^2\equiv M^2.
\ee
The absence of mixing terms is a consequence of the invariance of the
Lagrangian under the phase transformation $V^{\pm}_R\to\exp(\pm
i\alpha)V^{\pm}_R$. In fact, from (\ref{covariant}), only $V_{3R}$
mixes with the light vector fields. Notice that the parameter $r$ in
eq. (\ref{r-parameter}) can be written also in the following way
\be
r=\frac 1 4\frac {g^2 v^2}{M^2}.
\ee
The remaining two eigenvalues, in the limit of small  $r$ are (we
continue to call $W^\pm$, $V^{\pm}_L$  the mass eigenvectors)
\bea
M^2_{{W}}&=&\frac \vv 4 g^2 (1-r \sf^2 +\cdots),\nn\\ M^2_{{V}_L}&=&
\frac \vv 4 g^2 (\frac 1 r \frac 1 {\cf^2} +\frac {\sf^2}{\cf^2} +r
\sf^2 +\cdots).
\label{mw2}
\eea
Notice that for $r\to 0$, $M^2_{W}$ coincides with the  SM expression
for the $W$ mass.

Let us call {\bf C} the matrix which transforms the fields $(W^\pm,V_L^\pm)$
appearing in the
Lagrangian (\ref{3.1}) into the charged eigenstates.
At the first order in $r$ we get
\be
{\bf C}^{-1}=
\pmatrix{\cf(1-\sf^2 r) & -\sf (1+\cf^2 r)\cr
        \sf (1+\cf^2 r) & \cf(1-\sf^2 r) }.
\ee

\hfill\break\medskip\noindent
\underbar{Neutral gauge sector}
\hfill\break\medskip\noindent
In this sector there is a null eigenvector corresponding to the photon:
\be
\gamma = (s_{\tilde\theta} W_3 +c_{\tilde\theta} Y) \cpsi + \frac {1} {\sqrt{2} }
(V_{3L} + V_{3R})\spsi,
\ee
where
\bea
\tan {\tilde\theta} &=& {\cf \st} \sqrt{P},\nn\\
\tan\psi &=& \sqrt{2} s_{\tilde\theta} \frac {g_0} {g_2}=\sqrt{2}\frac{\sf\st}
{\sqrt{1- 2\sf^2\st^2}}.
\eea
The remaining eigenvalues are, again in the limit of small $r$,
\bea
M^2_{Z}&=&\frac \vv 4  \frac {g^2}{\ct^2}(1-r\sf^2
\frac {1-2\ct^2+2\ct^4}{\ct^4}
+\cdots),\nn\\ M^2_{V_{3L}}&=& \frac \vv 4 {g^2} ( \frac 1 {r\cf^2}
+\frac {\sf^2}{\cf^2}
-r\sf^2 \frac {\ct^2}{1-2\ct^2}+\cdots),\nn\\
M^2_{V_{3R}}&=&\frac \vv 4\frac {g^2}{\ct^2} (\frac 1{r} \frac {\dd
\ct^4}{ P} +\frac {\sf^2\st^4}{P} +r \frac {\dd\sf^2\st^8}{\dd \ct^4
(1-2\ct^2)}+\cdots).
\label{mz2}
\eea
Only for  $\varphi=0$ the heavy vectors are degenerate in mass.

Let us call {\bf N} the matrix which transforms the fields $(W_3,Y,V_{3L},V_{3R})$
appearing in the
Lagrangian (\ref{3.1}) into the neutral eigenstates which we will call
$(\gamma,Z,V_{3L},V_{3R})$. At the first order in $r$ we get
\be
{\bf N}^{-1}=
\pmatrix{\cf\st & \cf(\ct- \dd{\frac{\sf^2} \ct} r) & -\sf(1+\cf^2 r) & \dd{
\frac{ \cf \sf
       \st^4 \sqrt{P}}{\ct^3(1-2 \ct^2)}} r\cr
       \sqrt{P} & -\dd{\frac{\st}{\ct}} \sqrt{P} (1-\dd{\frac{\sf^2\st^2}{\ct^4}} r)&
-\dd{\frac{\cf\sf\st\sqrt{P}}{1-2 \ct^2}} r & -\dd{\frac{\sf\st}{\ct}}(1+\dd{\frac{
            \st^2 P}{\ct^4}}  r)\cr
\sf\st & \sf\ct(1+\dd{\frac{\cf^2}{\ct^2}} r )& \cf (1-\sf^2 r)&-\dd{
\frac{\st^2 P^{3/2}}
{\ct^3(1-2 \ct^2)}} r\cr
\sf\st &-\dd{\frac{\sf\st^2}{\ct}}(1+\dd{\frac{P}{\ct^4}} r)& \dd{\frac{\cf^3\st^2}
{1-2 \ct^2}} r & \dd{\frac{\sqrt{P}}\ct}(1-\dd{\frac{\sf^2
\st^4}{\ct^4}}r) }.
\ee

We will now consider the couplings of the vector bosons to the
fermions. We assume that the fermions have standard transformation
properties under the group $SU(2)_L\otimes U(1)_Y$, and therefore the
couplings to the heavy bosons arise only through the mixing.

In the charged sector, at the first order in $r$ the couplings are given by
\be
{\cal L}_{\rm fermions}^{\rm charged}
=-(h_W W_\mu^-+h_L V^{-}_{L\mu}) J_L^{\mu -}+h.c.,
\label{lch}
\ee
with
\be
h_W=\frac g {\sqrt {2}}(1-\sf^2 r),
\label{hw}
\ee
\be
h_L=-\frac g {\sqrt {2}}(1+\cf^2 r)\tan\varphi,
\label{hL}
\ee
and $J^\pm_L= \bar \psi_L \gamma^\mu \tau^\pm \psi_L$.
Notice that there is no coupling of $V_{R}^\pm$ to fermions, because
these particles do not mix with the $W^\pm$'s. Also, for $r=0$ the
couplings of $W^\pm$ to the fermions coincide with the standard ones.

In the neutral sector the couplings are defined by
\bea
{\LL}_{\rm fermions}^{\rm neutral}&=&-e J_{em} \gamma
 -[AJ_{3L}+B J_{em}]Z\nn\\
&&-[C J_{3L} +D  J_{em}]V_{3L}
-[E J_{3L} +F  J_{em}]V_{3R},
\label{lneu}
\eea
with
\be
e=g\st,
\label{ee}
\ee
 and
\bea
A&=&\frac g\ct (1-\sf^2 \frac {\st^4+\ct^4}{\ct^4}r),\nn\\ B&=& \frac
g\ct(-\st^2 +\frac {\sf^2\st^4}{\ct^4}r),\nn\\ C&=&\frac
g\ct(-\tan\varphi\ct +\frac {\cf\sf \ct^3}{2\ct^2-1}r),\nn\\ D&=&\frac
g\ct\frac {\cf\sf \st^2\ct}{2\ct^2-1}r,\nn\\ E&=&\frac
g\ct(\frac{\sf\st^2}{\dd\sqrt{P}}+\frac {\dd\sf\st^6\sqrt{P}}{\dd\ct
(1-2\ct^2)}r),\nn\\ F&=&\frac g\ct(-\frac{\sf\st^2}{\dd\sqrt{P}}-\frac
{\dd\sf\st^4\sqrt{P}} {\dd\ct^4}r).
\label{cou}
\eea
The expression for the electric charge is valid  to all order in $r$,
while the other coefficients in (\ref{cou}) are given only at first
order in $r$. In particular the couplings  of the $Z$ to fermions go
back to their SM values for $r\to 0$. Notice that  there are no
couplings increasing when $r\to 0$, both in the charged and in the
neutral sector.

We can rewrite the fermionic couplings of a generic gauge boson $V$ in
a form which will be useful later:
\be
\bar\psi  [v^V+a^V\gamma_5] \gamma_\mu\psi V^\mu,
\ee
where, by comparing with eqs. (\ref{lch}-\ref{cou}),
we get for the charged sector:
\bea
v^W &=& -\f{h_W}{2},~~~~~~~~~~~~~~a^W=v^W,\nn\\ v^{L} &=&
-\f{h_L}{2},~~~~~~~~~~~~~~a^{L}=v^{L},
\label{3.24}
\eea
and, for the neutral one:
\bea
v^Z &=& -(A\f{\tau_3}{4}+B
Q_{em}),~~~~~~~~~~~~~~a^Z=-A\f{\tau_3}{4},\nn\\ v^{3L} &=&
-(C\f{\tau_3}{4}+D Q_{em}),~~~~~~~~~~~~~~a^{3L}=-C\f{\tau_3}{4},\nn\\
v^{3R} &=& -(E\f{\tau_3}{4}+F
Q_{em}),~~~~~~~~~~~~~~a^{3R}=-E\f{\tau_3}{4}.
\label{3.25}
\eea

\resection{The $\epsilon$ parameters: tree level}

At tree level, the definition of the Fermi constant $G_F$ is
\be
\f {G_F}{\sqrt{2}}=\f 1 4 \left(\f {h_W^2}{M_W^2}+\f {h_L^2}{M_{V_L}^2}\right)
=\f 1{2v^2},
\label{gf}
\ee
This is an exact result, and it can be verified at the order $r$ by
using eqs. (\ref{hw}), (\ref{hL}) and (\ref{mw2}).

From the expression of $M_Z^2$ in
(\ref{mz2}), by using (\ref{ee}) and (\ref{gf}) we get at
the first order in $r$
\be
\ct^2=\ctz^2(1+r \Delta\f{\st^2}{\ct^2-\st^2}),
\label{costheta}
\ee
with
\bea
&&\Delta=\sf^2 \f{1-2 \ct^2+2 \ct^4}{\ct^4},\nn\\ &&\ctz^2=\f 1 2
+\sqrt{\f 1 4 -\f{\pi\alpha}{\sqrt{2} G_F M_Z^2}}.
\label{theta0}
\eea
Notice that the $\theta_0$ angle, here defined, coincides with the
$\theta$ angle given in eq. (44) of ref. \cite{degene}. In ref.
\cite{linear} we were interested in the leading order $r=0$ and so we
did not distinguish between $\theta$ and $\theta_0$.

By using the expressions for $M_W$  and $M_Z$ in eqs. (\ref{mw2}) and
(\ref{mz2}), the eq. (\ref{costheta}) and the definition of $\Delta
r_W$ given by
\be
\f{M_W^2}{M_Z^2}=\ctz^2 (1-\f{\st^2}{\ct^2-\st^2}\Delta r_W),
\ee
we obtain
\be
\Delta r_W=- r \f{\sf^2}{\ct^2}.
\ee
From the standard definition
\be
{\cal L}^{Z}_{\rm fermions}=-\f{e}{\stz\ctz}(1+\f{\Delta\rho}{2})
[J_{3L}-\stz^2(1+\Delta k)J_{em}],
\ee
by comparing with (\ref{lneu}) and using (\ref{gf}) we get
\bea
\Delta\rho &=& -r \sf^2 \f {\ct^4+\st^4}{\ct^4},\nn\\
\Delta k &=& -2 r \sf^2 \f{\st^2}{\ct^2-\st^2},
\eea
from which we extract the expressions for the $\eps$ parameters at the
first order in $r$ \cite{degene}:
\bea
\eps_1 &=& -r \sf^2\f{\ct^4+\st^4}{\ct^4},\nn\\
\eps_2 &=& -r \sf^2,\nn\\
\eps_3 &=& -r \f{\sf^2}{\ct^2}.
\label{eps}
\eea
This shows that at tree level the heavy sector decouples, at least as
far as its contribution to LEPI physics is concerned. The restrictions
on the parameter space coming from (\ref{eps}) have been recently
discussed in \cite{lhc}.

\resection{The $\epsilon$ parameters: one loop level}

To evaluate the radiative corrections to the $\eps$ parameters, we
will use the definition given in \cite{barbieri} which is more
suitable than the one used in the previous Section in terms of the
observables. Obviously the two definitions lead to the same result.
For instance the result in eq. (\ref{eps}) was verified in ref.
\cite{axial-vector} by using the procedure which will be discussed
below.

Following the definitions of the $\eps$ parameters given in \cite{barbieri}
we need to calculate, besides the corrections to the vacuum polarization
amplitudes for the $W$, $Z$ and $\gamma$,
 the contributions to the vector
and the axial-vector form factors at the $Z$ pole in the $Zl^+l^-$
vertex and the one loop corrections (boxes, vertices, new vector boson
 and fermion self-energies) to the $\mu$ decay amplitude
at zero external momenta.

Let us define the vacuum self-energies
\be
\Pi^{\mu\nu}_{ij}(p)=-i g^{\mu\nu}\Pi_{ij}(p^2)+p^\mu p^\nu~
{\rm terms},
\ee
where
\be
\Pi_{ij}(p^2)= A_{ij}(0)+p^2 F_{ij}(p^2),
\ee
with $i,j=W,~\gamma,~Z$.

The corrections to the vector and axial-vector form factors at $p^2=M_Z^2$
in the $Z$ leptonic interactions from proper vertex and fermion self-energies
are parameterized as
\be
-i\f e {2 \ctz\stz}\bar v \gamma_\mu[\delta g_V-\gamma_5 \delta g_A] u,
\label{dgv}\ee
where $\theta_0$ is defined in eq. (\ref{theta0}).

The third contribution comes from the one loop corrections to $G_F$
from the $\mu$ decay (except the $W$ self-energy \cite{barbieri}):
\be
-i~ \delta G_F~ [\bar e \gamma_\mu(1-\gamma_5) \nu_e][\bar \nu_\mu \gamma^\mu
(1-\gamma_5) \mu].\label{dgf}
\ee
In terms of these quantities one can express the $\epsilon$ parameters
as
\bea
\eps_1&=&e_1-e_5 -\f {\delta G_F}{G_F}-4\delta g_A,\nn\\
\eps_2&=&e_2-\stz^2 e_4 -\ctz^2 e_5-\f {\delta G_F}{G_F}-\delta g_V
-3\delta g_A,\nn\\
\eps_3&=&e_3+\ctz^2 e_4 -\ctz^2 e_5+\f {\ctz^2-\stz^2}{2\stz^2} \delta g_V
-\f {1+2\stz^2}{2\stz^2} \delta g_A,
\eea
with
\bea
e_1&=&\f { A_{33}(0)-A_{WW}(0)}{M^2_W},\nn\\
e_2&=&F_{WW}(M^2_W)-F_{33}(M^2_Z),\nn\\ e_3&=&\f {\ctz}{\stz}
F_{30}(M^2_Z),\nn\\
e_4&=&F_{\gamma\gamma}(0)-F_{\gamma\gamma}(M^2_Z),\nn\\ e_5&=&M^2_Z
F^\prime_{ZZ}(M^2_Z),
\label{ei}
\eea
where the indices $0,3$ refer to $Y,~ W_3$ bosons and the following relations
hold:
\bea
\Pi_{30}&=&-\stz\ctz \Pi_{ZZ}+ \stz\ctz \Pi_{\gamma\gamma}+
(\ctz^2-\stz^2) \Pi_{Z\gamma},\nn\\
\Pi_{33}&=& \ctz^2 \Pi_{ZZ}+ 2\stz\ctz \Pi_{Z\gamma}+\stz^2 \Pi_{\gamma\gamma}.
\eea

We will evaluate the contribution to the $\eps$ parameters at one loop
level, by using  dimensional regularization  for the UV divergences in
the loops and we will introduce  the arbitrary mass scale parameter
$\mu$. Since  we are interested in the decoupling properties of the
model only for observable quantities, the ultraviolet divergent terms
will not play any role and in fact they cancel out. Therefore we will
not perform the full renormalization procedure of the model.

\resection{Vacuum polarization amplitudes}
We list here the results for the vector boson self-energy diagrams. In
evaluating  the vacuum polarization amplitudes, since we are
interested in proving the decoupling, we keep only the potentially
dangerous terms and we neglect terms proportional to $r$. Of course we
will not consider diagrams with only light particles because they give
the SM contribution plus corrections of order $r$ that we neglect. All
the relevant couplings are given in the Appendix. We will list here
the contributions to the various vacuum polarization functions.

For the $W$ self-energy, we have contributions from the graphs
$S_1$, $S_2$ and $S_3$ (Fig. 1). In particular, by indicating with
$\Pi^{(i)}$ the amplitude from
$S_i$, we get
\bea
\Pi_{WW}^{(1)}(p)&=&g^2\Big \{
r^2 \sf^2\cf^2 A_1(p,M_W,M_{V_{3L}})+r^2 \f {\sf^2 \st^4 P}{\ct^6}
 A_1(p,M_W,M_{V_{3R}})\nn\\
&+&(1-2 r (1-2 \cf^2)) A_1(p,M_{V_{L}},M_{V_{3L}})+ r^2 \f {\sf^2
\cf^2}{\ct^2} A_1(p,M_{V_{L}},M_Z)\Big\},
\eea
\be
\Pi_{WW}^{(2)}(p)=\f {g^2} 2 (1-2 r (1-2 \cf^2))
\Big\{ A_2(M_{V_{L}})+A_2(M_{V_{3L}})\Big\},
\ee
\be
\Pi_{WW}^{(3)}(p)=g^2 \f{\sf^2}{\cf^2} M_W^2
A_3(p,M_{\rho_{U}},M_{V_{L}})+ 2 g^2 \f r {\sf^2}q^2 M_W^2
A_3(p,M_{\rho_R},M_W).
\ee
The functions $A_i$ are the result of the various loop integrals. As
already said, we will neglect all the contributions to the
self-energies going to zero with $r$. For this reason we give the
explicit expressions of the $A_i$ functions only up to the order which
leads to non vanishing results. The exact results for $A_1$ and $A_3$
can be found in \cite{grazzini}. We have
\be
A_1(p,M_H,M_H)= \f 1 {16\pi^2} \Big\{(\f 9 2 M_H^2+7 p^2) Y_{H}
 -\f 3 4 M_H^2-\f 2 3 p^2\Big\}+ {\cal O}(\f 1 {M_H^2}),
\label{A10}
\ee
\be
A_1(p,m,M_H)=\f 1 {16\pi^2}
\f {M_H^4}{m^2} X_H
+ {\cal O}(M_H^2),
\label{A1}
\ee
\be
A_2(M_H)=-\f 1 {16\pi^2} \Big\{
\f 9 2  M_H^2 Y_H-\f 3 4 M_H^2 \Big\},
\label{A2}
\ee
\be
A_3(p,m,M_H)=\f 1 {16\pi^2} X_H + {\cal O}(\f 1 {M_H^2}),
\label{A3}
\ee
\be
A_3(p,M_H,m)=\f 1 {16\pi^2} \f {M_H^2}{m^2}\left(\f 1 4 Y_H-\f 3
8\right) + {\cal O}(\log{M_H^2}),
\ee
with
\bea
Y_H&=&-\f 2 \eps +\gamma +\log \f {M_H^2}{4\pi \mu^2},\nn\\ X_H&=&\f 5
8 -\f 3 4 Y_H,
\eea
where $\epsilon=4-D$, with $D$ the space-time dimension,  $\gamma$
the Euler constant and $M_H$  the mass increasing with $M$. Notice
that in the $\Pi^{(1)}_{WW}(p)$ term, there is no contribution from
the $(V_L,V_{3R})$ exchange, because the coupling $W V_L V_{3R}$ is of
order $r$ (see eq. (\ref{A.8})) and the loop contribution is ${\cal
O}(M^2)$.

Let us comment on the Higgs particle exchange in the $S_3$ loops.
Since the first  loop contribution is ${\cal O}(\log (M))$, only the
constant part in $r$ of ${\cal L}^h_{{\rm heavy-light}}$ given in eq.
(\ref{A.2}) is relevant. As a consequence we have to consider only the
$\rho_U$ exchange (see Fig. 1), since the $\rho_L$ and $\rho_R$
exchanges are suppressed by a $\sqrt{r}$ factor in the vertex.  There
is also a contribution from the $W\rho_R$ exchange (see Fig. 1),
because the loop diagram is now behaving as $M^2\log(M)$, and so the
factors $\sqrt{r}$ in the vertices are not enough to suppress this
term. However, this is true only for the momentum constant part in the
self-energy. Then, it can be seen immediately that there is no
correction to the $e_1$ parameter, due to the custodial symmetry. That
is this contribution cancels with the analogous one coming from
$\Pi_{ZZ}$ (see Fig. 2).

Notice that  $M_{V_L}$ and $M_{V_{3L}}$ differ of terms of order $r$,
therefore they can be taken to be equal at the order we consider here.
Their common value will be called $M_{V_L}$.

Summing all the contributions and retaining only the leading order in
$r$ we get
\bea
\Pi_{WW}(p) &=& \f{g^2}{16\pi^2}\Big\{
7 p^2 Y_{V_L} + 3 M^2_W \f {\sf^2}{\cf^2} X_{V_L} + M_W^2
\f{\sf^2 \st^4}{\ct^2 P}
 X_{V_{3R}}-\f 2 3 p^2\nn\\
 &+&\f {2q^2}{\sf^2}M_W^2\f{M_{\rho_R}^2}{M^2}
 \left(\f 1 4 Y_{\rho_R}-\f 3 8\right)
 \Big\}.
\eea

For the $Z$ self-energy, we have contributions from the graphs $S_1$, $S_2$
and $S_3$ (Fig. 2):
\bea
\Pi_{ZZ}^{(1)}(p)&=&g^2\Big\{
2 r^2 \f {\sf^2 \cf^2}{\ct^2}
A_1(p,M_W,M_{V_{L}})+
(\ct^2+2 r (2 \cf^2-1)) A_1(p,M_{V_{L}},M_{V_{L}})\nn\\ &+&
\f {\st^4} {\ct^2} (1+\f {2 r }{\ct^4}P) A_1(p,M_{V_{R}},M_{V_{R}})\Big\},
\eea

\be
\Pi_{ZZ}^{(2)}(p)=g^2\Big\{(\ct^2+2 r (2 \cf^2-1))
  A_2(M_{V_{L}})+\f {\st^4} {\ct^2} (1+\f {2 r }{\ct^4}P) A_2(M_{V_{R}})
\Big\},
\ee

\bea
\Pi_{ZZ}^{(3)}(p)&=&g^2\Big\{\f{\sf^2}{\cf^2\ct^2} M_W^2
A_3(p,M_{\rho_{U}},M_{V_{3L}})+
\f{\sf^2\st^4}{\ct^4 P} M_W^2
A_3(p,M_{\rho_{U}},M_{V_{3R}})\nn\\ &+& 2 \f r{\ct^2\sf^2}q^2M_Z^2
A_3(p,M_{\rho_R},M_Z)\Big\}.
\eea
As far as the Higgs particles exchange is concerned,
the same comment  we have done for  $\Pi_{WW}^{(3)}$ holds.

Summing up all the contributions and using eqs.
(\ref{A10})-(\ref{A3}), we get
\bea
\Pi_{ZZ}(p) &=& \f{g^2}{16\pi^2}\Big\{
7 p^2 (\ct^2 Y_{V_L} +\f {\st^4}{\ct^2 } Y_{V_R})
+ 3 M^2_Z \f {\sf^2}{\cf^2} X_{V_L} + M_Z^2 \f{\sf^2 \st^4}{\ct^2 P}
 X_{V_{3R}}-\f 2 3 p^2\ct^2 (1+\f {\st^4}{\ct^4})\nn\\
 &+&2\f{q^2}{\ct^2\sf^2}M_W^2 \f{M_{\rho_R}^2}{M^2}\left(\f 1 4
 Y_{\rho_R} - \f 3 8\right) \Big\}.
\eea

The contributions to the photon self-energy come
from the graphs $S_1$ and $S_2$ (Fig. 3):
\be
\Pi_{\gamma\gamma}^{(1)}(p)=g^2\st^2 \Big\{A_1(p,M_{V_{L}},M_{V_{L}})+
 A_1(p,M_{V_{R}},M_{V_{R}})\Big\},
\ee

\be
\Pi_{\gamma\gamma}^{(2)}(p)=g^2 \st^2 [ A_2(M_{V_{L}})+ A_2(M_{V_{R}})].
\ee
By using eqs. (\ref{A10}), (\ref{A2}), we get
\be
\Pi_{\gamma\gamma}(p) =
 p^2 \f {g^2}{16\pi^2}\st^2(7  (Y_{V_L} + Y_{V_R})-\f 4 3 ).
\ee

To the $\gamma Z$ self-energy contribute  the graphs $S_1$ and $S_2$ (Fig. 4):
\be
\Pi_{\gamma Z}^{(1)}(p)=g^2\Big\{\st \ct (1-\f r {\ct^2}(1-2 \cf^2))
A_1(p,M_{V_{L}},M_{V_{L}})-
 \f {\st^3}{\ct}(1+\f r {\ct^4}P) A_1(p,M_{V_{R}},M_{V_{R}})\Big\},
\ee
\be
\Pi_{\gamma Z}^{(2)}(p)=g^2\Big\{\st \ct (1-\f r {\ct^2}(1-2 \cf^2))
A_2(M_{V_{L}})-  \f {\st^3}{\ct}(1+\f r {\ct^4}P)
 A_2(M_{V_{R}})\Big\}.
\ee
Again by using eqs. (\ref{A10}), (\ref{A2}), we get
\be
\Pi_{\gamma Z}(p) =
 p^2\f{g^2}{16\pi^2} [7(\st\ct Y_{V_L} - \f {\st^3}{\ct} Y_{V_R})
+\f 2 3  \st \ct (\f {\st^2}{\ct^2}-1)].
\ee

From eq. (\ref{ei}) and the previous results we obtain (up to
corrections ${\cal O} (r)$)
\be
e_1=e_2=e_3=0.
\ee
Furthermore $e_4$ and $e_5$ are zero at the order here considered, because
$F_{\gamma\gamma}$ and $F_{ZZ}$ are independent of $p^2$.

\resection{Expressions for the relevant loops}

The loop diagrams  which will be relevant for the calculation, besides
the gauge boson self-energy loops  studied in the previous Section,
are listed in Fig. 5. Here we will give the generic expression for
these loops in the $M\to
\infty$ limit. We have explicitly verified that, doing this limit in the
loop integrand of the amplitudes, no
singularity appears in the integration over the Feynman parameters.
So we can safely  expand  the  amplitudes in $1/M$.

For the graph (a) the amplitude is given by
\def\slash#1{\ooalign{$\hfil/\hfil$\crcr$#1$}}
\bea
&&g_{V_1 V_2 V_3}~\eps_{abc}\int\f{d^Dk_1}{(2\pi)^D}~((2p+k_1)^\nu
g^{\mu\rho}+(k_1-p)^\rho g^{\mu \nu}-(2k_1+p)^\mu g^{\nu \rho})\nn\\
&&~~~~~~~~~~~~(-i)\left(g_{\nu\alpha}-\f{k_{1\nu}
k_{1\alpha}}{M^2_1}\right) (-i)\left(g_{\rho\sigma}-\f{(p+k_1)_{\rho}
(p+k_1)_{\sigma}}{M^2_2}
\right)\nn\\
&&~~~~~~~~~~~~i(v^1+a^1\gamma_5)\gamma^\alpha
i\f{\slash{k}+\slash{k}_1} {(k+k_1)^2}
i(v^2+a^2\gamma_5)\gamma^\sigma
\f{1}{k_1^2-M_1^2}\f{1}{(p+k_1)^2-M_2^2},
\label{7.1}
\eea
where $p$ is the momentum of the incoming gauge boson $V_3$,
$k$ and $p-k$ are the momenta of the outgoing fermions, which we will
take  massless. In eq. (\ref{7.1})
 $g_{V_1 V_2 V_3}$ is the trilinear gauge coupling which can be
directly read from eqs. (\ref{A.7}-\ref{A.9}), $v^i$ and $a^i$
$(i=1,2)$ are the vector and axial vector couplings of the gauge
vector bosons $V^i$ to the fermions (see eqs. (\ref{3.24}-\ref{3.25}))
and  $M_{1(2)}$ is the mass of the gauge boson $V_{1(2)}$.

In the $M_1>>M_2$ limit, the term $[(p+k_1)_\rho (p+k_1)_\sigma/M_2^2]
(g_{\nu\alpha}-
k_{1\nu} k_{1\alpha}/M_1^2)$ gives the   leading contribution:
\be
g_{V_1 V_2 V_3}~\eps_{abc} \left\{\Big (
[v^1 v^2+a^1 a^2] \gamma_\mu+[v^1 a^2+a^1 v^2] \gamma_5\gamma_\mu \Big )
\f 1 {M_1^2} A_1(p, M_2,M_1)+{\cal O}(\log M_1)\right\},\label{7.2}
\ee
where $A_1$ is given in eq. (\ref{A1}). We have considered just the
leading behavior because this  loop  always contributes to the
amplitudes with a suppression factor ${\cal O}(r)$ as we will see
explicitly in the following.

For large $M_1=M_2$, eq. (\ref{7.1}) would give terms ${\cal O}(\log
M_1)$, but cancellations, due to  mass degeneracy, occur and therefore
 one gets a finite result:
\be
g_{V_1 V_2 V_3}~\eps_{abc}\left\{\Big (
[v^1 v^2+a^1 a^2] \gamma_\mu+[v^1 a^2+a^1 v^2] \gamma_5\gamma_\mu \Big)
\f 3 2 \f 1 {16 \pi^2}+{\cal O}(\f 1 {M_1^2})\right\}.\label{7.3}
\ee

\def\vup{{{v^1}^\prime}}
\def\aup{{{a^1}^\prime}}
\def\vdp{{{v^2}^\prime}}
\def\adp{{{a^2}^\prime}}

For the graph (b), in Fig. 5, the amplitude is given by:
\bea
\int\f{d^Dk_1}{(2\pi)^D}&& i(v^{1^\prime}+a^{1^\prime} \gamma_5)\gamma^\nu
i(\slash{p}+\slash{k}_1)i(v^3+a^3\gamma_5)\gamma^\mu
i\slash{k}_1
i(v^1+a^1\gamma_5)\gamma^\rho\nn\\
&&(-i)\left(g_{\nu\rho}-\f{(k+k_1)_\nu(k+k_1)_\rho}{M_1^2}\right)
\f{1}{(k+k_1)^2-M_1^2}\f{1}{(p+k_1)^2}\f{1}{k_1^2},\label{7.4}
\eea
where $p$ is the momentum of the incoming  gauge boson $V_3$, $k$ and
$p-k$ are the momenta of the outgoing fermions again considered
massless and we have used the $prime$ to distinguish between the two
vertices of the $V_1$ gauge boson with the fermions. Here $v^3$ and
$a^3$ are the couplings of the external gauge boson $V_3$ to the
fermion pair. Since the amplitude is dimensionless, in the large $M_1$
limit, one expects terms which are at most ${\cal O}(\log M_1)$.
However the explicit calculation shows again that these divergent
terms cancel out leaving a finite contribution:
\bea
&&-\Big\{
\Big[v^3[v^1 \vup +a^1 \aup]+a^3[v^1 \aup+a^1 \vup]\Big]\gamma_\mu\nn\\
&&+\Big[v^3[v^1 \aup+a^1 \vup]+a^3[v^1 \vup+a^1 \aup]\Big]\gamma_5\gamma_\mu
\Big\}
\f 3 2 \f i {16 \pi^2}+{\cal O}(\f 1 {M_1^2}).\label{7.5}
\eea

For the graph (c),
the amplitude is given by:
\bea
\int\f{d^Dk_1}{(2\pi)^D}&& i(v^{1^\prime}+a^{1^\prime} \gamma_5)\gamma^\nu
i(\slash{p}-\slash{k}_1)
i(v^1+a^1\gamma_5)\gamma^\rho\nn\\
&&(-i)\left(g_{\nu\rho}-\f{{k_1}_\nu {k_1}_\rho}{M_1^2}\right)
\f{1}{k_1^2-M_1^2}\f{1}{(p-k_1)^2}.\label{7.6}
\eea
 In the large $M_1$ limit, one expects terms which are at most
${\cal O}(\log M_1)$. However the explicit
calculation shows that these divergent terms cancel out leaving a finite
contribution:
\be
-\f{i}{16 \pi^2}{\slash {p}} \f 3 2 \Big\{{(v^1)}^2
+{(a^1)}^2-2 v^1 a^1 \gamma_5+{\cal O}(\f 1 {M_1^2})\Big\}.\label{7.7}
\ee
Notice that we have also neglected  UV divergent terms of the type
$\slash {p} (p^2/M_1^2) (2/\eps)$. These terms, which arise in theory
with massive vector fields in the unitary gauge, are not a problem for
the renormalizability since one can show \cite{collins} that all the
corresponding counterterms vanish by the equations of motion. Or, said
in different words, they do not contribute to the $S$-matrix elements.

For the graph (d) we find:
\bea
\int\f{d^Dk}{(2\pi)^D}&&
\Big\{[i(v^2+a^2\gamma_5)\gamma^\rho i(\slash{p}_1-\slash{k})
i(v^1+a^1\gamma_5)\gamma^\mu]\nn\\
&&\otimes[i(v^{2^\prime}+a^{2^\prime}\gamma_5)\gamma^\sigma
i(-\slash{p}_2+\slash{k})i(v^{1^\prime}+a^{1^\prime}\gamma_5)\gamma^\nu]\nn\\
&&(-i)\left(g_{\mu\nu}-\f{k_\mu k_\nu}{M^2_1}\right)
(-i)\left(g_{\rho\sigma}-\f{(p_1-k+p_1^\prime)_\rho(p_1-k+p_1^\prime)_\sigma}
{M^2_2}\right)\Big\}\nn\\
&&\f{1}{k^2-M^2_1}\f{1}{(p_1-k)^2}
\f{1}{(p_1-k+p_1^\prime)^2-M^2_2}\f{1}{(k-p_2)^2}.
\label{pippo}
\eea
The momenta of the external fermions can be read in Fig. 5-(d) and
 we have used the $prime$ to denote the  couplings of the gauge bosons
$V_1$ and $V_2$ to the fermion pairs on the right-hand-side of the
figure.

In the case $M_1>>M_2$, the leading contribution comes from
 terms of the type $k^4/M_2^2$ and $k^6/(M_2^2 M_1^2)$ in the
expression between curl brackets in eq. (\ref{pippo}). The result is:
\bea
&&
i\Big\{ (v^1 v^2 +a^1 a^2)(\vup \vdp+\aup \adp)\gamma_\mu\otimes \gamma^\mu\nn\\
   &&+(v^2 a^1 +a^2 v^1)(\vdp \aup+\adp \vup)\gamma_5 \gamma_\mu \otimes
        \gamma_5\gamma^\mu\nn\\
&&+ (v^2 v^1 +a^2 a^1)(\vdp \aup+\adp \vup)\gamma_\mu \otimes
     \gamma_5\gamma^\mu\nn\\
&&+ (v^2 a^1 +a^2 v^1)(\vdp \vup+\adp \aup)\gamma_5\gamma_\mu \otimes
\gamma^\mu\Big\}\nn\\
&&\times
 [\f 1 {M_1^4} A_1(p,M_2,M_1)+{\cal O}(\f 1 {M_1^2})].
\label{box}
\eea
In the calculation of the four fermion amplitude we have to take into
account, in addition to the box diagram of  Fig. 5-(d), the one with
the exchange $V_1\to V_2$ and the corresponding crossed diagrams. In
the limit $M_1>>M_2$  it turns out that the amplitude corresponding to
the  $V_1\to V_2$ exchange is still given by eq. (\ref{box}) while the
result for each of crossed boxes give minus the amplitude of eq.
(\ref{box}). Finally let us observe that the exchange of two heavy
particles in the box diagrams is suppressed by an additional $1/M^2$
factor and therefore it will be neglected.

Notice that, apart from the finite terms, all the relevant amplitudes
in the limit are expressed through the  same function $A_1$, which
occurs in the calculation of the vector boson self-energy corrections.

\resection{One loop corrections to $G_F$}

The one loop corrections to $G_F$ from the $\mu$ decay (except the $W$
self-energy \cite{barbieri}) are illustrated in Fig. 6. The black dots
stand for the proper vertices, fermion
 and new vector  boson self-energy contributions, as
illustrated in Figs. 8-12.

First of all let us consider the box diagrams. The relevant contributions are
given in Fig. 7 where for simplicity we have not drawn the crossed diagrams.
Due to the fact that for the $\mu$ decay process we necessarily have
the exchange of a charged gauge boson, whose coupling to fermions satisfy
the relation $v=a$, the box diagram contribution of eq. (\ref{box})
can be rewritten
as
\be
i v^1 \vup( v^2 +a^2)(\vdp+\adp)
\f 1 {M_1^4} A_1(p,M_2,M_1)
 [\gamma_\mu+\gamma_5\gamma_\mu]\otimes [\gamma^\mu+\gamma_5\gamma^\mu],
\ee
where $v_1$ and $v_1'$ denote the couplings of the charged gauge boson
$V_1$ to the fermion pairs. So the one loop corrections to $G_F$
coming from the box diagrams are of the form of eq. (\ref{dgf}).

In particular it is easy to show that, due to the couplings of
$V_{3R}$ to fermions, the sum of the four amplitudes, corresponding to
the direct and the crossed box diagrams,  of the $(W,V_{3R})$ exchange
vanishes. The sum of the contributions from the $(W,V_{3L})$ and
$(V_L,Z)$ exchanges (Fig. 7) is (taking into account also the crossed
diagrams):
\be
-i\delta G_F^{(a)}=-i \f {g^4} 8 \f {\sf^2}{\cf^2}
\f 1 {M^4_{V_{3}}} \Big[A_1(p,M_W,M_{V_{3L}})+
\ct^2 A_1(p,M_Z,M_{V_{L}})\Big].
\ee
We have neglected the box diagrams with $(V_L,\rho_U)$ exchange since the $\rho_U$
couplings to fermions are proportional to their masses.

To compute the Fig. 6-(b) amplitudes we need the $W$ vertex
corrections, $\Gamma^\mu_{We\bar \nu}$, given in Fig. 8 and the $V_L$
ones, $\Gamma^\mu_{V_L e\bar \nu}$, given in Fig. 9. Because these are
corrections to charged gauge boson vertices, the results given in eqs.
(\ref{7.2}), (\ref{7.3}), (\ref{7.5}) factor out in  terms
proportional to $\gamma_\mu+\gamma_5\gamma_\mu$.

To the $\Gamma^\mu_{We\bar \nu}$ contribute two types of loops $L_1$ and $L_2$.
For each graph $L_1$ in Fig. 8 we have also to consider the one obtained by
exchanging the external fermionic lines. In particular in the case of the
$(W,V_{3R})$ exchange the total contribution vanishes due to the $V_{3R}$
fermion couplings.
The result coming from the $L_1$ loops is the following:
\be
\Gamma_{W e{\bar \nu}}^{\mu(1)}= - i  \f {g^3}{2\sqrt 2}
\f{ \sf^2}{\cf^2}  \Big[ \f {r \cf^2} {M^2_{V_L}} \Big(A_1(p, M_Z, M_{V_L})
+A_1(p, M_W, M_{V_{3L}})\Big)-\f 3 2\f 1 {16 \pi^2}
\Big](\gamma^\mu+\gamma_5\gamma^\mu),
\ee
where the last term is due to the $(V_L,V_{3L})$ exchange (see eq. (\ref{7.3})).

The result from the $L_2$ loop is:
\be
\Gamma_{W e{\bar \nu}}^{\mu(2)}= -\f i{16 \pi^2}  \f 3 2 \f {g^3}{8\sqrt 2}
\f {\sf^2}{\cf^2}\Big[ 1-\f {\cf^2 \st^4}{\ct^2 P}\Big]
(\gamma^\mu+\gamma_5\gamma^\mu).
\ee

To the $\Gamma^\mu_{V_L e\bar \nu}$ contributes only the $L_1$ loop.
This is because in the four fermion amplitude there is an additional
factor $1/M^2$ coming from the $V_L$ propagator. As already said,
for each graph $L_1$ in Fig. 9 we have also to consider the one obtained by
exchanging the external fermionic lines.
The result is the following:
\be
\Gamma_{V_L e{\bar \nu}}^{\mu}= -  i  \f {g^3}{2\sqrt 2}
\f {\sf}{\cf} \f {1}{M^2_{V_L}} \Big[\ct^2 A_1(p, M_Z, M_{V_{L}})+
A_1(p, M_W, M_{V_{3L}})\Big](\gamma^\mu+\gamma_5\gamma^\mu).
\ee

Let us now evaluate the contribution of the fermion self-energies to
$\Gamma^{\mu}_{V_3 f f^{\prime}}$ where $V_3$ is a generic vector boson (Fig. 10).
Of course the self-energy insertion on the external fermionic legs must be taken
 with a factor 1/2.
The explicit expression for each of the graphs of Fig. 10,
in the $M_1\to\infty$ limit, is, by using eq. (\ref{7.7})
\be
\Gamma^{\mu (s.e.)}
_{V_3 f f^{\prime}}=\f {3i} {32\pi^2}\f 1 2 \Big\{
\big[v^3((v^1)^2+(a^1)^2)+2 a^3 v^1 a^1\big]\gamma^\mu
+\big[a^3((v^1)^2+(a^1)^2)+2 v^3 v^1
a^1\big]\gamma_5\gamma^\mu\Big\}.\label{8.5}
\ee
Again, when we consider a charged boson vertex correction the result factors out
in a term proportional to $\gamma^\mu+\gamma_5\gamma^\mu$.
Using the general expression (\ref{8.5}) we can evaluate the
contribution to $\Gamma_{W e{\bar \nu}}^{\mu}$ due to the
self-energy corrections coming from  the exchange
of $V_L$, $V_{3L}$ and $V_{3R}$:
\be
\Gamma_{W e{\bar \nu}}^{\mu(s.e.)}= -\f i {16 \pi^2} \f 3 2 \f {g^3}{8\sqrt 2}
\f {\sf^2}{\cf^2}\Big[ 3 +\f {\cf^2 \st^4}{\ct^2 P}\Big]
(\gamma^\mu+\gamma_5\gamma^\mu).
\ee
%So the total result for $\Gamma_{W e{\bar \nu}}^{\mu}$ is
%\be
%\Gamma_{W e{\bar \nu}}^{\mu}= - i  \f {g^3}{2\sqrt 2}
%\f {\sf^2}{\cf^2}\Big[ \f {M^2_W}{M^4_{V_L}} \Big(A_1(p, M_Z, M_{V_L})
%+A_1(p, M_W, M_{V_{3L}})\Big)\Big](\gamma^\mu+\gamma_5\gamma^\mu)
%\ee

The sum of the contributions coming from the graphs in Fig. 6-(b,c)
is again of the form given in eq. (\ref{dgf}) with
\be
-i \delta G_F^{(b,c)}=-i
 \f {g^4}4 \f {\sf^2}{\cf^2}
\f 1 {M^4_{V_{L}}}  A_1(p,M_Z,M_{V_{L}})\st^2.
\ee

The contribution to the four fermion amplitudes due to the $WV_L$
self-energy  (Fig. 6-(d)) comes from the loops in Fig. 11. In the case
of loops with two heavy gauge bosons, as we have already found in the
calculation of the vector boson self energies, there is a cancellation
of the most divergent terms with the corresponding tadpole
contributions. As a result, the sum of the loops $S_1$ and $S_2$ with
two heavy gauge bosons is ${\cal O}(\log M)$ and so it is suppressed
in the four fermion amplitude due to the factor $1/M^2$ coming from
the $V_L$ propagator. For the same reason we have not drawn loops with
only a logarithmic divergence, like for example the one with the
$(V_L,\rho_U)$ exchange.

The only graphs giving a finite contribution in the $M\to\infty$ limit
are the $S_1$ ones with the $(W,V_{3L})$ and $(V_{3L},Z)$ exchanges
(Fig. 11). The result is again in the form of eq. (\ref{dgf}) with
\be
-i\delta G_F^{(d)}= i \f {g^4} 4 \f {\sf^2}{\cf^2}
\f 1 {M^4_{V_{L}}}  \Big[ A_1(p,M_W,M_{V_{3L}})+  A_1(p,M_Z,M_{V_{L}}) \Big].
\ee

Finally, let us consider
the contribution from the $V_L$ vacuum polarization diagrams (Fig. 6-(e))
given in Fig. 12 where we have considered only loops
which diverge at least as $M^4$.
The corresponding correction to $G_F$ is
\be
-i\delta G_F^{(e)}= -i\f {g^4} 8 \f {\sf^2}{\cf^2}
\f 1 {M^4_{V_{L}}}  \Big[ A_1(p,M_W,M_{V_{3L}})+
\ct^2 A_1(p,M_Z,M_{V_{L}})\Big].
\ee
Notice that  contribution to the $V_L$ self-energy coming from the
loop with one $V_L$ and one heavy Higgs boson, $\rho_{L,R}$, is
vanishing in the limit, although the corresponding trilinear couplings
are increasing with $\sqrt{r}$. In fact, the loop is only logarithmic
in $M$. The sum of the $\delta G_F^{(a,b,c,d,e)}$ contributions
vanishes.

\resection{One loop corrections to the $Ze^+e^-$ vertex}

To evaluate the extra contributions  to the vector and axial-vector form factors
$\delta g_V$ and $\delta g_A$ at the $Z$ pole we need, besides the
one loop vertex corrections, the fermion self-energies
and the "heavy-light"
vector boson self-energies in which the light boson is a $Z$ (Fig. 13).

The one loop contribution to $\Gamma_{Z e^+ e^-}^\mu$ given in Fig.
13-(a) is the sum of the graphs $L_1$ and $L_2$ (Fig. 14). In
particular from the graph $L_1$ we get:
\be
\Gamma_{Z e^+ e^-}^{\mu (1)} = \left[i \f {g^3} 2 \f{\sf^2}{\ct} r
\f 1 {M^2_{V_L}} A_1(p,M_W,M_{V_L})+\f {g^3} 4 \ct \f{\sf^2}{\cf^2}
(-\f 3 2 \f i{16\pi^2})\right] (\gamma^\mu+\gamma_5\gamma^\mu).
\ee
The first term comes from two equal contributions from
the $(V_L^+,W^-)$  and $(V_L^-,W^+)$ exchanges, the second one
from the $(V_L,V_L)$ exchange.

To the graph $L_2$ we have three contributions from the $V_{3L}$,
$V_{L}$ and $V_{3R}$ exchanges:
\bea
\Gamma_{Z e^+ e^-}^{\mu (2)} &=&\Big\{ \Big\{ \left[
\f{g^3}{16}  \f{\sf^2}{\cf^2\ct}(1 -2 \st^2)
\right]+\left[
-\f{g^3}{8 }  \f{\sf^2}{\cf^2\ct}
\right]\Big\}(\gamma^\mu+
\gamma_5\gamma^\mu)\nn\\
&&+
\f{g^3}{16 } \f {\sf^2\st^4}{\ct^3 P}\left[(1 -10 \st^2)\gamma_\mu+
(1+6 \st^2) \gamma_5\gamma^\mu\right]
\Big\} (-\f 3 2 \f i {16\pi^2}).
\eea
The  amplitude corresponding to Fig. 13-(b) get  contributions from
the graphs $S_1$ and $S_2$ in Fig. 15. In the case of loops with two
heavy gauge bosons, as we already said, there is a cancellation of the
most divergent terms with the corresponding tadpole contributions. As
a result, the sum of $S_1$ and $S_2$ in this case is ${\cal O}(1 /
M^2)$ due to the heavy gauge boson propagator. So the only non
vanishing term in the amplitude of Fig. 13-(b), comes from two equal
contributions from the $(V_L^+,W^-)$  and $(V_L^-,W^+)$ exchanges in
the loop. The result is:
\be
\Gamma_{Z e^+ e^-}^{\mu (b)} = -i \f {g^3} 2 \f{\sf^2}{\ct} r
\f 1 {M^2_{V_L}} A_1(p,M_W,M_{V_L})(\gamma^\mu+\gamma_5\gamma^\mu).
\ee

Concerning the contribution from Fig. 13-(c) the relevant $ZV_{3R}$
self-energy diagrams are given in Fig. 16. Since we have the same
cancellation of the most divergent terms between the graphs $S_1$ and
$S_2$ the result is ${\cal O}(1/ M^2)$ due to the heavy gauge boson
$V_{3R}$ propagator.

Finally, the corrections to the $Ze^+e^-$ vertex due to the fermionic
self-energy contributions (Fig. 13-(d)) come from the exchange of
$V_{3L}$, $V_{L}$ and $V_{3R}$. As already observed for the $\delta
G_F$ calculation, these terms must be considered with a factor $1/2$,
and using eq. (\ref{8.5}) we get
\bea
\Gamma_{Z e^+ e^-}^{\mu (d)} &=&
\Big\{\left[
\f{g^3}{16}  \f{\sf^2}{\cf^2\ct}
3 (1 -2 \st^2)
\right](\gamma^\mu+\gamma_5\gamma^\mu)\nn\\
&&+\left[
\f{g^3}{16 } \f {\sf^2\st^4}{\ct^3 P}(1 -10 \st^2)\gamma_\mu+
(1+6 \st^2) \gamma_5\gamma^\mu\right]
\Big\}(\f 3 2 \f i {16\pi^2}).
\eea
The sum of  all the one loop corrections  to the $Ze^+e^-$ vertex
vanishes at the leading order in the $M\to\infty$ limit. Therefore,
from the definition given in eq. (\ref {dgv}) we get
\be
\delta g_V=\delta g_A=0.
\ee

We have also checked that  in this limit we have no extra corrections
at the leading order to the vertices $\Gamma_{Z f_1 f_2}^{\mu}$, and
$\Gamma_{W f_1 f_2}^{\mu}$.

\resection{Conclusions}

We have developed an extension of the SM based on the gauge group
$SU(2)_L\otimes U(1)\otimes SU(2)_L'\otimes SU(2)_R'$ with two
different energy scales: the electroweak one, $v$, and a higher scale
$u$. The model is a linear realization of a dynamical breaking of the
electroweak symmetry previously proposed, containing two new triplets
of spin one particles $V_L$ and $V_R$ (degenerate BESS). The interest
in this model was due to its decoupling property: in the limit of
infinite mass of the heavy vector bosons ($u\to\infty$), one recovers
the Higgsless SM.
In the linear version one has also scalar states in the spectrum, and
in this case, in the limit of large $u$, one gets back to the SM.

To
show the decoupling we have considered the observables relevant to
LEPI physics. In particular we have computed the tree value of the
$\epsilon$ parameters, which turns out to be ${\cal O} (v^2/u^2)$.
Being the Lagrangian of the model renormalizable, we have also shown
that the decoupling property holds also at the level of radiative
corrections.

We have performed the calculation of the contributions to the
$\epsilon$ parameters due to the new physics at one loop by evaluating
the self-energy corrections to $W,~Z,~\gamma$ propagators, the vertex
corrections to $Z e^+e^-$ and to the Fermi coupling constant.
Dimensional regularization and unitary gauge have been used. The
result is a cancellation of all divergent and finite contributions in
the $u\to\infty$ limit. The corrections to the $\epsilon$ observables
at the order $v^2/u^2$ turn out to be, in general, numerically smaller
than the tree level ones (taken at the same order), due to the factor
$1/16\pi^2$ coming from the loops.

Even if we are not proving in general the complete decoupling of the
high-energy sector, we have shown that at the LEPI energy the model is
undistinguishable from the SM, whereas the signatures at high energy
can be very different \cite{lhc}.
\newpage

\renewcommand{\theequation}{A.\arabic{equation}}
\appsection

From eq.
(\ref{3.1}), using the new couplings defined in eq. (\ref{2.10}),
and expressing the
vector and Higgs fields in terms of the corresponding mass eigenstates,
we derive the Higgs-vector interactions
at the leading order in $r$.

Let us observe that, since there are trilinear couplings of the order
$1/\sqrt{r}$
(see eq. (\ref{3.1})), and we have evaluated the mass diagonalization
matrices up to
${\cal O}(r)$, the result for ${\cal L}^h$ is correct up to ${\cal O}(\sqrt{r})$.
For the light sector we obtain
 an expression which coincides with the analogous
one in the SM (remember that now $W$, $V_L$, $V_R$, $Z$, $V_{3L}$,
$V_{3R}$ as well as $\rho_U,~ \rho_L,~ \rho_R$ denote the mass
eigenstates)
\be
{\cal L}^h_{\rm light}=\frac {g^2} 4
(\rho_U^2+2\rho_Uv)(W^+W^-+
\frac 1 {2\ct^2}Z^2).
\ee
%For the heavy-light sector we get
%\bea
%{\cal L}^h_{\rm heavy-light}&=&\frac {g^2} 4
%\Big\{(\rho_U^2+2\rho_Uv)[-\tan\varphi(W^+V^-_L+W^-V^+_L+\frac 1\ct
%ZV_{3L})\nn\\
%&+&
%\frac{\sf\tan^2\theta}{\sqrt{P}}ZV_{3R}+
%\tan^2\varphi~( V^+_LV^-_L+\f 1 2 V_{3L}^2)
%\nn\\
%&+& \f 1 2 \f {\sf^2 \st^4}{\ct^2 P} V_{3R}^2-\f{\sf^2\st^2}{\cf\ct\sqrt{P}}V_{3L}
%V_{3R}]\nn\\
%&+& 2 v \sqrt{r}
%[\frac{1}{\cf}(W^+V^-_L+W^-V^+_L+\f{1}{\ct} Z V_{3L})\rho_L\nn\\
%&-&\f{\st^2}{\ct^2 \sqrt{P}} Z V_{3R}\rho_R]\Big\}
%\eea
%and the heavy sector we get
%\bea
%{\cal L}^h_{\rm heavy}&=&\frac {g^2} 4
%\Big\{ \f{1}{\cf^2 \sf^2}(\rho_L^2+2\f{\sf}{\sqrt{r}} v \rho_L) (V_L^+V^-_L
%+\f 1 2 V_{3L}^2)\nn\\
%&+&\frac 1{\sf^2}(\rho_R^2+2 \f{\sf}{\sqrt{r}} v \rho_R) (V_R^+V_R^-
%+\f 1 2 \f {\ct^2}{P} V_{3R}^2)
%\nn\\
%&-& \f{2 v \sqrt{r}\st^2}{\sf (1-2 \ct^2)} (\f{\sqrt{P}}{\cf\ct} \rho_L+
%\f{\cf\ct}{\sqrt{P}} \rho_R)V_{3L} V_{3R}
%\Big\}
%\eea

For the heavy-light sector we get
\bea
{\cal L}^h_{\rm heavy-light}&=&\frac {g^2} 4
\Big\{(\rho_U^2+2\rho_Uv+ \f 2\sf \sqrt{2 r} q \rho_U\rho_R)
[-\tan\varphi(W^+V^-_L+W^-V^+_L+\frac 1\ct
ZV_{3L})\nn\\
&+&
\frac{\sf\tan^2\theta}{\sqrt{P}}ZV_{3R}+
\tan^2\varphi~( V^+_LV^-_L+\f 1 2 V_{3L}^2)
\nn\\
&+& \f 1 2 \f {\sf^2 \st^4}{\ct^2 P} V_{3R}^2-\f{\sf^2\st^2}
{\cf\ct\sqrt{P}}V_{3L}
V_{3R}]\nn\\
&+& \f 2 \sf q v \sqrt{2 r}\rho_R
[-\tan\varphi(W^+V^-_L+W^-V^+_L+\frac 1\ct
ZV_{3L}) +
\frac{\sf\tan^2\theta}{\sqrt{P}}ZV_{3R}]\nn\\
&+& \f 2 \sf \sqrt{2r} q  \rho_R (\rho_U+v)(W^+W^-+\frac 1 {2\ct^2}Z^2)\nn\\
&+&  v \sqrt{ 2 r}
[\frac{1}{\cf}(W^+V^-_L+W^-V^+_L+\f{1}{\ct} Z V_{3L})(\rho_L+\rho_R)\nn\\
&-&\f{\st^2}{\ct^2 \sqrt{ P}} Z V_{3R}(\rho_R-\rho_L)]\nn\\
&-&
 \f{1}{\cf^2 \sf^2}\rho_U q ( \f{\sqrt{2 r}}{\sf}(\rho_L+\rho_R)+2 v )
 (V_L^+V^-_L
+\f 1 2 V_{3L}^2)\nn\\
&+&\frac 1{\sf^2}\rho_U q (\f {\sqrt{2 r}}{\sf} (\rho_L-\rho_R)-2 v) (V_R^+V_R^-
+\f 1 2 \f {\ct^2}{P} V_{3R}^2)
\Big\}, \label{A.2}
\eea
and for the heavy sector we get
\bea
{\cal L}^h_{\rm heavy}&=&\frac {g^2} 4
\Big\{ \f{1}{\cf^2 \sf^2}[\f 1 2 (\rho_L+\rho_R)^2+\sqrt{2}\f{\sf}{\sqrt{r}}
v (\rho_L+
\rho_R)-\f {\sqrt{2r}}{\sf} v q^2\rho_R)] (V_L^+V^-_L
+\f 1 2 V_{3L}^2)\nn\\
&+&\frac 1{\sf^2}
[\f 1 2 (\rho_L-\rho_R)^2-\sqrt{2}\f{\sf}{\sqrt{r}} v (\rho_L-
\rho_R)-\f {\sqrt{2r}}{\sf} v q^2\rho_R)]
 (V_R^+V_R^-
+\f 1 2 \f {\ct^2}{P} V_{3R}^2)
\nn\\
&+& 2 \rho_R \sqrt{2r} v \f q {\sf}
[\tan^2\varphi~( V^+_LV^-_L+\f 1 2 V_{3L}^2)
+ \f 1 2 \f {\sf^2 \st^4}{\ct^2 P} V_{3R}^2-\f{\sf^2\st^2}{\cf\ct\sqrt{P}}V_{3L}
V_{3R}]\nn\\
&-& \f{v \sqrt{2 r}\st^2}{\sf (1-2 \ct^2)} (\f{\sqrt{P}}{\cf\ct} (\rho_L+\rho_R)+
\f{\cf\ct}{\sqrt{P}} (\rho_R-\rho_L))V_{3L} V_{3R}
\Big\}.
\eea

Concerning the Higgs self interactions, we  can obtain the scalar
potential in terms of the Higgs field eigenstates by simply rewriting
eq. (\ref{pot}) in terms of the transformed fields.

Finally, let  us derive the vector boson self-couplings. Notice that,
since in ${\cal L}^{kin}$ given in eq. (\ref{lkin}) the couplings are
${\cal O} (1)$ and we have evaluated the mass diagonalization matrices
up to ${\cal O} (r)$, the result for ${\cal L}^{kin}$ is correct up to
${\cal O} (r)$.

Let us define the following formal combination
\be
AB^-C^+=A^{\mu\nu}B_\mu^-C_\nu^++A^\nu(B_{\mu\nu}^-C^{\mu +}-
B_{\mu\nu}^+C^{\mu -}),
\ee
where
\be
A^{\mu\nu}=\partial^\mu A^\nu-\partial^\nu A^\mu,
\ee
and similar expression for $B_{\mu\nu}^\pm$. Then
the trilinear gauge boson couplings in terms of the original fields are
given by
\be
\LL^{\rm tril} =i[g_0 W_3 W^- W^+
+g_2 V_{3L} V^{-}_{L} V^+_{L} +g_2 V_{3R} V^{-}_{R} V^+_{R}].
\ee
Using the redefinition of the couplings and the expressions for the mass
eigenstates, we find, again at the first order in $r$, for the
light sector
\be
{\cal L}_{\rm light}^{\rm tril}=ig[\st \gamma W^-W^++\ct Z W^-W^+],
\label{A.7}
\ee
for the heavy-light sector
\bea
{\cal L}_{\rm heavy-light}^{\rm tril}&=&ig [\st \gamma ({V}_L^-{V}_ L^+ +
{V}_R^-{V}_ R^+)
+(\ct +r\frac 1 \ct (2
\cf^2-1))Z {V}_L^-{V}_L^+\nn\\
 &+&\cf \sf \frac
{r}{\ct}(ZW^-{V}_L^++Z{V}_L^- W^+)-\frac {\st^2}{ \ct} (1+r \frac {1}
{\ct^4} P)Z{V}_R^-{V}_R^+\nn\\ &+&(1-r(1-2\cf^2))({V}_{3L} W^-{V}_L^+
+{V}_{3L}{V}_L^-W^+) +r\cf\sf  {V}_{3L}W^- W^+\nn\\
&-&r\frac{\cf\st^2\sqrt{P}}{\ct(1-2\ct^2)}({V}_{3R}W^-{V}_L^+ +
{V}_{3R}{V}_L^-W^+)\nn\\ &+&r\frac{\sf\st^2\sqrt{P}}{\ct^3}{V}_{3R}W^-
W^+],
\label{A.8}
\eea
and for the heavy sector
\bea
{\cal L}_{\rm heavy}^{\rm tril}&=&ig [ (\f {2 \cf^2-1}{\cf\sf} -3 \cf\sf r)
V_{3L} {V}_ L^- {V}_L^+
+\f{\cf^3\st^2}{\sf(1-2 \ct^2)} r V_{3L} {V}_R^-{V}_R^+\nn\\
 &+& \f{\st^2(\sf^2-\ct^2)\sqrt{P}}{ \sf \ct^3(1-2 \ct^2)} r V_{3R}{V}_L^-V_L^+
+\f{\sqrt{P}}{\sf\ct}(1-r \f{\sf^2\st^4}{\ct^4}){V}_{3R}{V}_R^-V_R^+].
\label{A.9}
\eea

The quadrilinear couplings are obtained starting from
\bea
&\LL^{\rm quad}=
-{\dd \frac { g_0^2} 2} S_{\mu\nu\rho\sigma}&
[W_\mu^+ W_\nu^- ( W_\rho^+ W_\sigma^- + W_{3\rho} W_{3\sigma})\nn\\
&&+\frac 1 {\tan^2\varphi} V_{L\mu}^+ V_{L\nu}^- ( V_{L\rho}^+
V_{L\sigma}^- + V_{3L\rho} V_{3L\sigma})\nn\\ &&+\frac 1
{\tan^2\varphi} V_{R\mu}^+ V_{R\nu}^- ( V_{R\rho}^+ V_{R\sigma}^- +
V_{3R\rho} V_{3R\sigma})],
\eea
with $S_{\mu\nu\rho\sigma}=2g_{\mu\nu} g_{\rho\sigma}
-g_{\mu\rho} g_{\nu\sigma} -g_{\mu\sigma} g_{\nu\rho}$.

At the lowest order in $r$
one gets for the light part
\bea
{\cal L}_{\rm light}^{\rm quad}=&- {\dd\frac {  g^2}  2}
S^{\mu\nu\rho\sigma} & [W_\mu^+ W_\nu^- ( W_\rho^+ W_\sigma^-+\ct^2
Z_\rho Z_\sigma\nn\\ &&+2\ct\st \gamma_\rho Z_\sigma +\st^2
\gamma_\rho\gamma_\sigma)],
\eea
for the heavy-light part
\bea
{\cal L}_{\rm heavy-light}^{\rm quad}=&-{\dd \frac {  g^2}  2}
S^{\mu\nu\rho\sigma} &
\{ (1- 2 r(1-2 \cf^2)) W_\mu^+ W_\nu^- (
{V}_{3L\rho}{V}_{3L\sigma}+{V}_{L\rho}^+ {V}_{L\sigma}^-)\nn\\
&&+(W_\mu^+{V}_{L\nu}^-+{V}_{L\mu}^+ W_\nu^-) [(1-2 r (1-2 \cf^2))
({V}_{L\rho}^+ W_\sigma^- + W_\rho^+ {V}_{L\sigma}^-)\nn\\
&&+(\frac{2\cf^2-1}{\cf\sf}+ r \f{1-6 \cf^2+6 \cf^4}{\cf\sf})
({V}_{L\rho}^+{V}_{L\sigma}^- + {V}_{3L\rho}{V}_{3L\sigma})]\nn\\
&&+{V}_{L\mu}^+ {V}_{L\nu}^- [(1-2 r (1-2\cf^2)) W_\rho^+ W_\sigma^- +
(\ct^2 -2 r (1-2\cf^2)) Z_\rho Z_\sigma\nn\\ &&+2(\ct\st-r
\f{\st}{\ct}(1-2 \cf^2)) \gamma_\rho Z_\sigma + \st^2
\gamma_\rho\gamma_\sigma\nn\\ &&+(\frac{2\cf^2-1}{\cf\sf}+r \f{1-6
\cf^2+6 \cf^4}{\cf\sf}) ({V}_{L\rho}^+W_{\sigma}^-+
{W}_{\rho}^+{V}_{L\sigma}^-)\nn\\ &&+2(\f{2 \cf^2 -1}{\cf\sf}\ct+r
\f{1-3 \cf^2\sf^2(1+\ct^2)}{\cf\sf\st})
 {V}_{3L\rho}{Z}_{\sigma}\nn\\
&&+2 (\f{2 \cf^2 -1}{\cf\sf}\st-3 r \cf \sf\st) {V}_{3L\rho}
{\gamma}_{\sigma}]\nn\\
&&+{V}_{R\mu}^+{V}_{R\nu}^-[(\frac{\st^4}{\ct^2}+2 r \f{\st^4}{\ct^6}
P) Z_\rho Z_\sigma- 2(\frac{\st^3}{\ct}+r \f{\st^3}{\ct^5} P)
\gamma_\rho Z_\sigma\nn\\
&&+\st^2\gamma_\rho\gamma_\sigma
-2\frac{\st^2\sqrt{P}}{\sf\ct^2}(1+r \f{-2\sf^2\st^2+\ct^2(1+\sf^2\st^2)}
{\ct^4}) {V}_{3R\rho}{Z}_{\sigma}\nn\\ &&+2\f{\st \sqrt{P}}{\ct\sf}
(1-r \f{\sf^2\st^4}{\ct^4}){V}_{3R\rho}{\gamma}_{\sigma}]\},
\eea
and for the heavy part
\bea
{\cal L}_{\rm heavy}^{\rm quad}=&-{\dd \frac {  g^2}  2} S^{\mu\nu\rho\sigma} &
\{
(\f{1-3 \cf^2+3 \cf^4}{\cf^2\sf^2}+4 r (1-2 \cf^2))\nn\\ && V_{L\mu}^+
V_{L\nu}^- ( {V}_{3L\rho}{V}_{3L\sigma}+{V}_{L\rho}^+
{V}_{L\sigma}^-)\nn\\ &&+ V_{R\mu}^+ V_{R\nu}^- [\f{P}{\sf^2\ct^2}(1-2
r \f{\sf^2\st^4}{\ct^4}) {V}_{3R\rho}{V}_{3R\sigma}\nn\\
&&+\f{1}{\sf^2}{V}_{R\rho}^+ {V}_{R\sigma}^-]\}.
\eea

Both the trilinear and quadrilinear light parts of the Lagrangian
agree with the SM results, and the heavy-light sectors do not show any
coupling increasing with the heavy mass $M$.

\par\noindent
\vspace*{1cm}
\par\noindent
{\bf Acknowledgements}
\par\noindent
This work has been carried out within the Program Human Capital and
Mobility: ``Tests of electroweak symmetry breaking and future European
colliders'', CHRXCT94/0579.

\newpage
\begin{center}
\Large{
{\bf Figure Captions} }
\end{center}

\noindent
{\bf Fig. 1} - Graphs contributing to the $W$ self-energy,
{$\Pi_{WW}$}.

\noindent
{\bf Fig. 2} - Graphs contributing to the $Z$ self-energy,
{$\Pi_{ZZ}$}.

\noindent
{\bf Fig. 3} - Graphs contributing to the $\gamma$ self-energy,
{$\Pi_{\gamma\gamma}$}.

\noindent
{\bf Fig. 4} - Graphs contributing to the $\gamma Z$ self-energy,
{$\Pi_{\gamma Z}$}.

\noindent
{\bf Fig. 5} - The generic loop diagrams for the evaluation of the
$\epsilon$ parameters (except for the vector boson self-energies).

\noindent
{\bf Fig. 6} - One loop diagrams for the $\mu$-decay, necessary to
evaluate the corrections to $G_F$.

\noindent
{\bf Fig. 7} - The box diagrams relevant to the corrections to $G_F$.

\noindent
{\bf Fig. 8} - The vertices {$\Gamma^\mu_{We\bar \nu}$} relevant to
the corrections to $G_F$.

\noindent
{\bf Fig. 9} - The vertices {$\Gamma^\mu_{V_L e\bar \nu}$} relevant to
the corrections to $G_F$.

\noindent
{\bf Fig. 10} - The fermion self-energy  contributions to the generic
vertex {$\Gamma^{\mu (s.e.)}_{V_3 f f^{\prime}}$} relevant to the
corrections to $G_F$.

\noindent
{\bf Fig. 11} - Graphs contributing to the $WV_L$ self-energy,
{$\Pi_{W V_L}$}.

\noindent
{\bf Fig. 12} - Graphs contributing to the $V_L$ self-energy,
{$\Pi_{V_LV_L}$}.

\noindent
{\bf Fig. 13} - One loop  diagrams relevant to the corrections to
$\delta g_V$, $\delta g_A$.

\noindent
{\bf Fig. 14} - The vertices {$\Gamma_{Z e^+ e^-}^{\mu}$} relevant to
the corrections to $\delta g_V$, $\delta g_A$.

\noindent
{\bf Fig. 15} - Graphs contributing to the $ZV_{3L}$ self-energy,
{$\Pi_{Z V_{3L}}$}.

\noindent
{\bf Fig. 16} -   Graphs contributing to the $ZV_{3R}$ self-energy,
{$\Pi_{Z V_{3R}}$}.

\newpage
\begin{figure}
\epsfysize=10truecm
\centerline{\epsffile{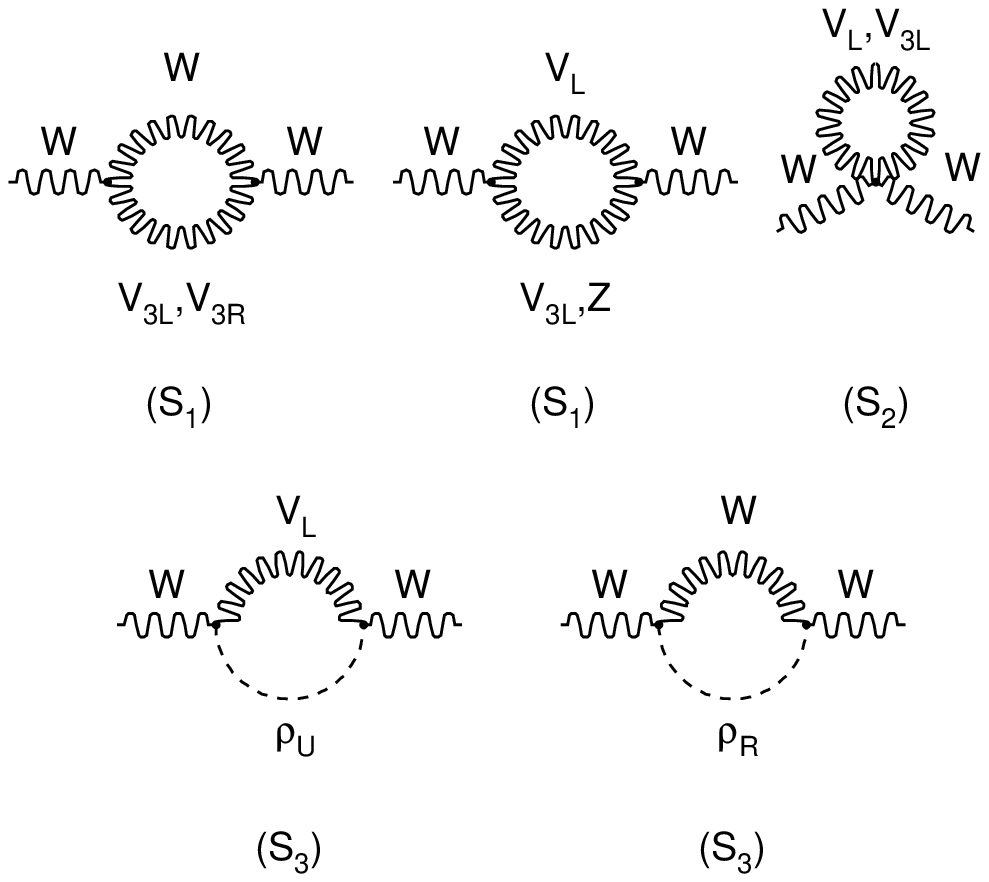}}
\noindent
\centerline{\bf Fig. 1}
\end{figure}

\begin{figure}
\epsfysize=10truecm
\centerline{\epsffile{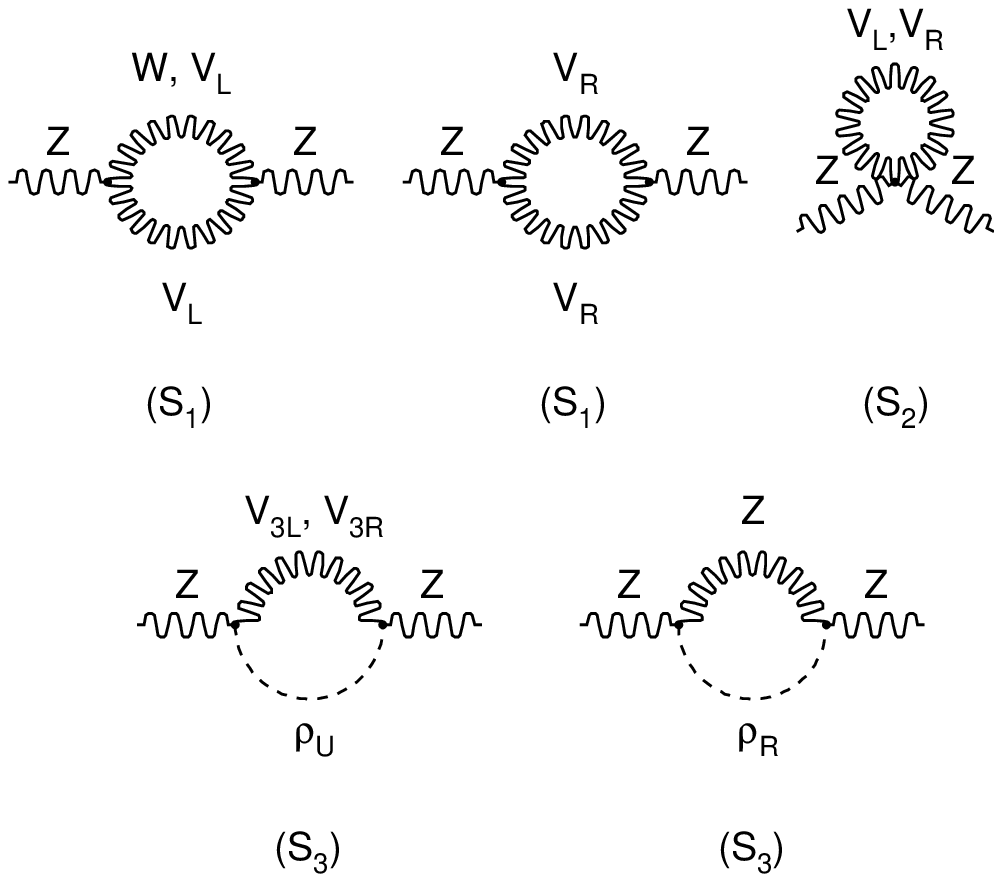}}
\noindent
\centerline{\bf Fig. 2}
\end{figure}

\begin{figure}
\epsfysize=6truecm
\centerline{\epsffile{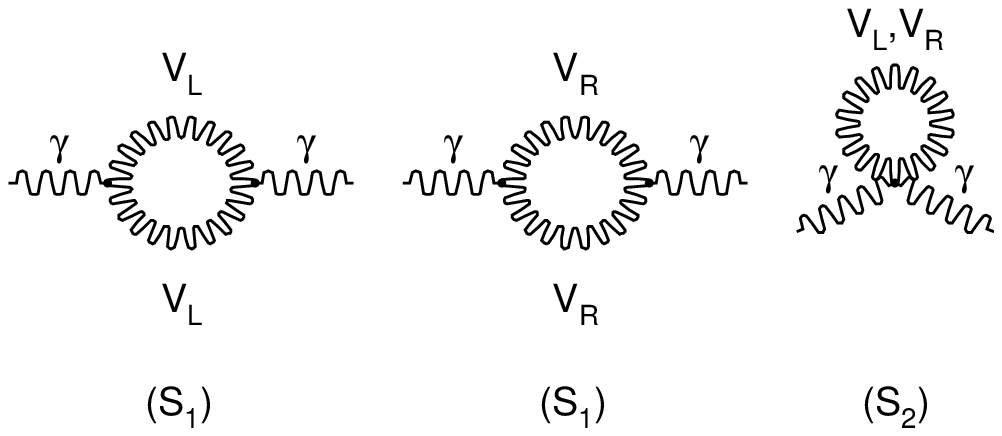}}
\noindent
\centerline{\bf Fig. 3}
\end{figure}

\begin{figure}
\epsfysize=6truecm
\centerline{\epsffile{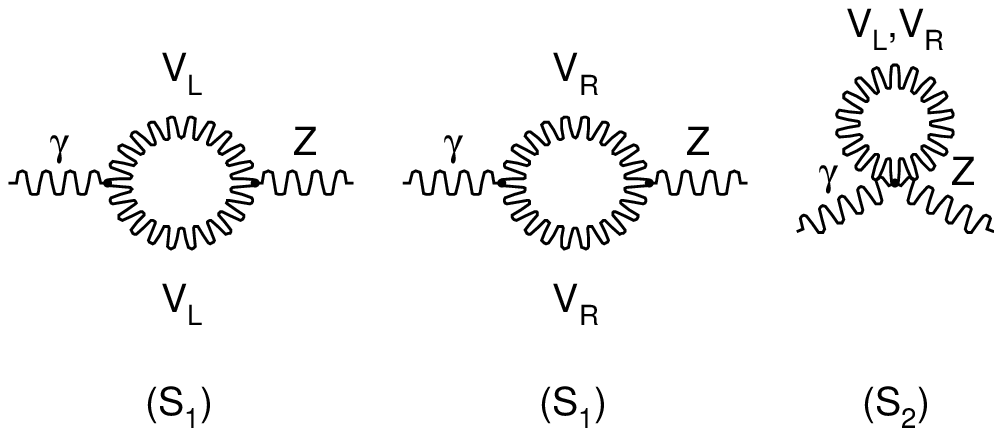}}
\noindent
\centerline{\bf Fig. 4}
\end{figure}

\begin{figure}
\epsfysize=10truecm
\centerline{\epsffile{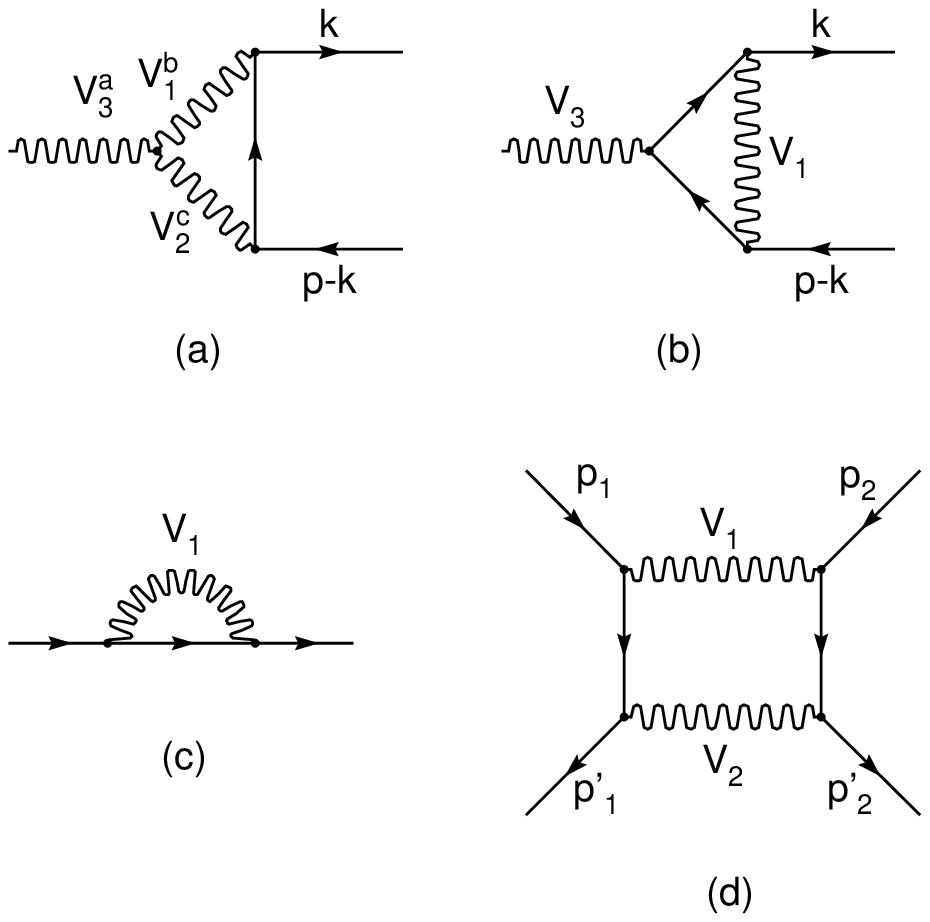}}
\noindent
\centerline{\bf Fig. 5}
\end{figure}

\begin{figure}
\epsfysize=12truecm
\centerline{\epsffile{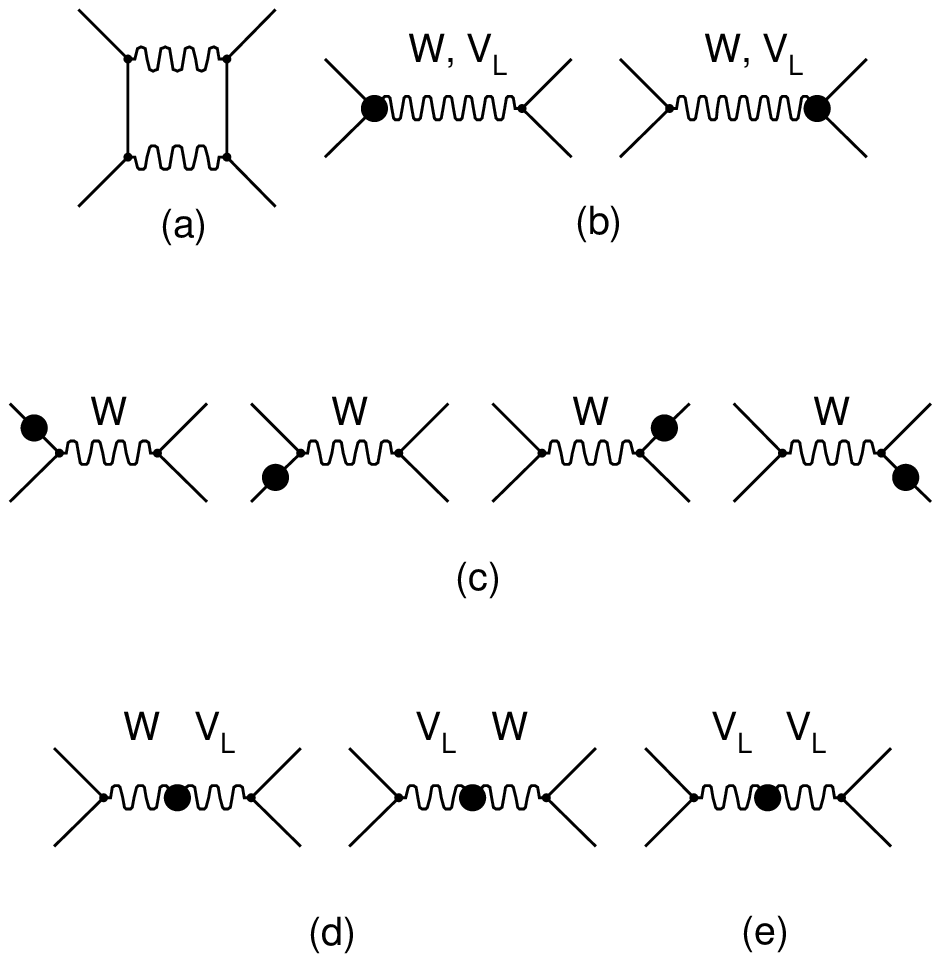}}
\noindent
\centerline{\bf Fig. 6}
\end{figure}

\begin{figure}
\epsfysize=4truecm
\centerline{\epsffile{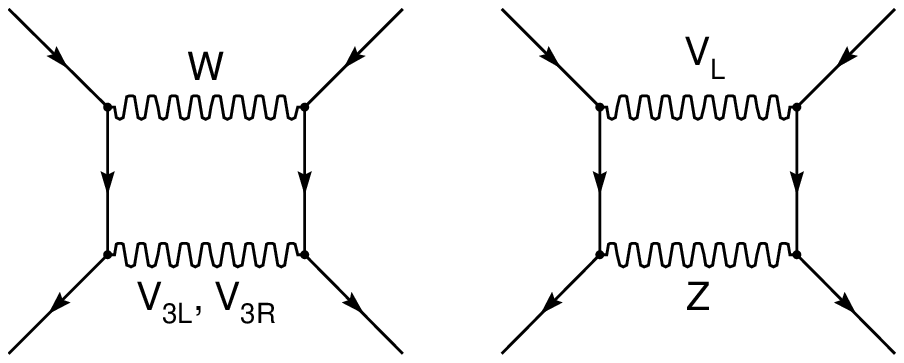}}
\noindent
\centerline{\bf Fig. 7}
\end{figure}

\begin{figure}
\epsfysize=9truecm
\centerline{\epsffile{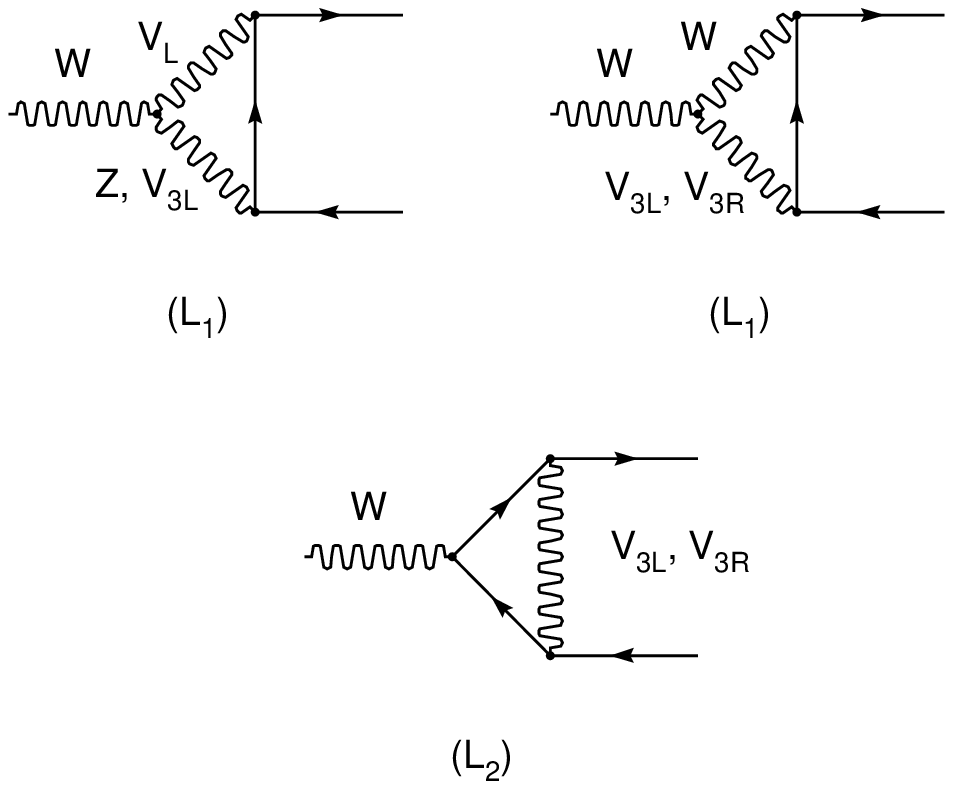}}
\noindent
\centerline{\bf Fig. 8}
\end{figure}

\begin{figure}
\epsfysize=6truecm
\centerline{\epsffile{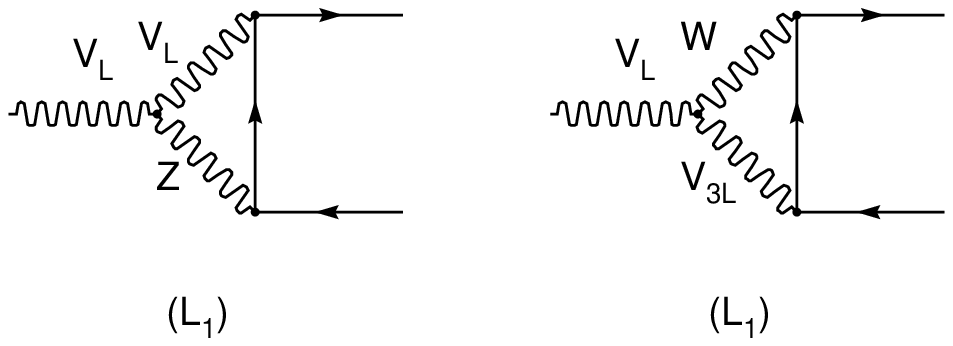}}
\noindent
\centerline{\bf Fig. 9}
\end{figure}

\begin{figure}
\epsfysize=6truecm
\centerline{\epsffile{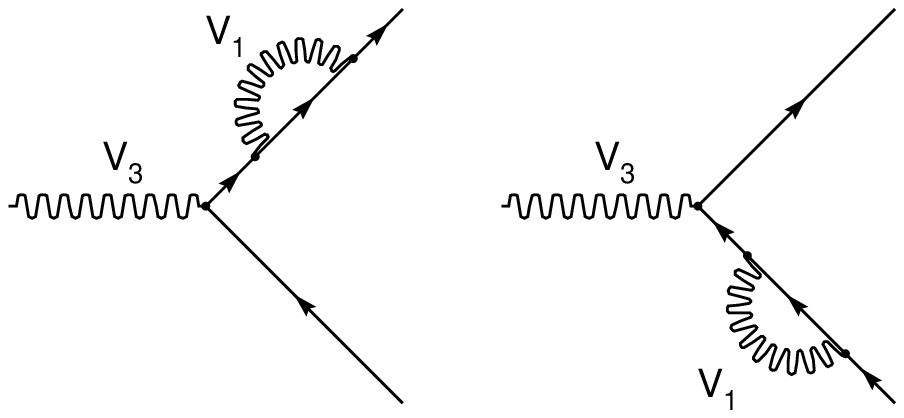}}
\noindent
\centerline{\bf Fig. 10}
\end{figure}

\begin{figure}
\epsfysize=6truecm
\centerline{\epsffile{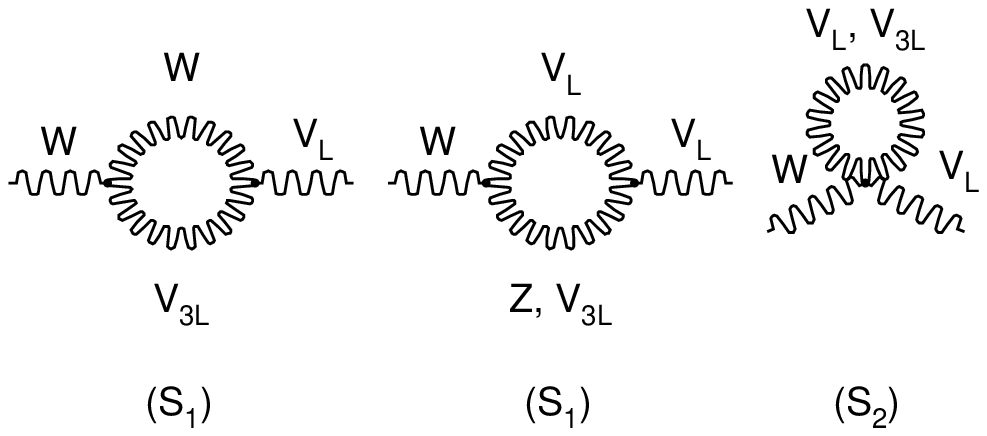}}
\noindent
\centerline{\bf Fig. 11}
\end{figure}

\begin{figure}
\epsfysize=6truecm
\centerline{\epsffile{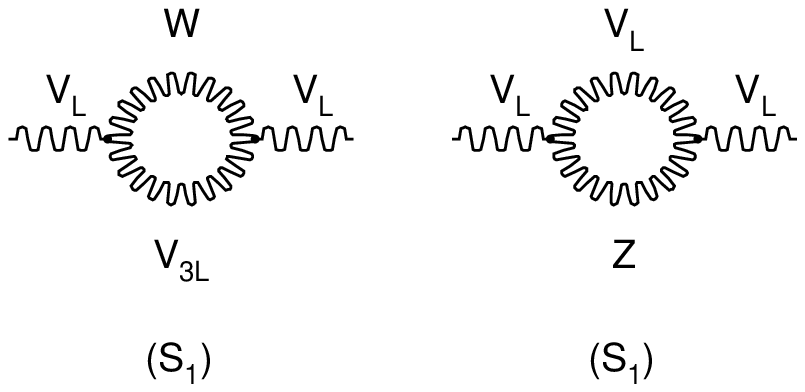}}
\noindent
\centerline{\bf Fig. 12}
\end{figure}

\begin{figure}
\epsfysize=9truecm
\centerline{\epsffile{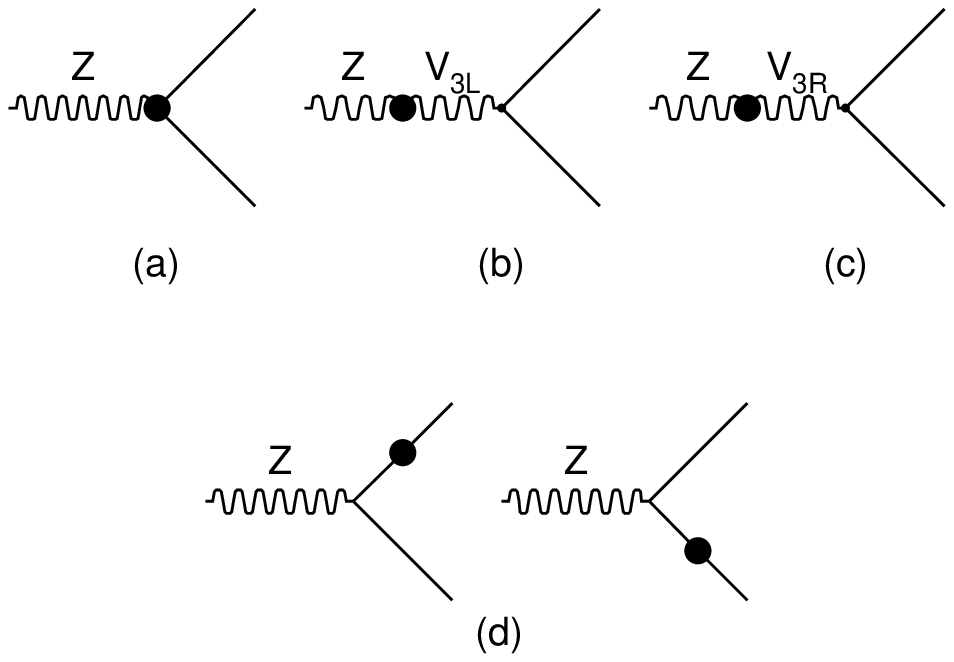}}
\noindent
\centerline{\bf Fig. 13}
\end{figure}

\begin{figure}
\epsfysize=5truecm
\centerline{\epsffile{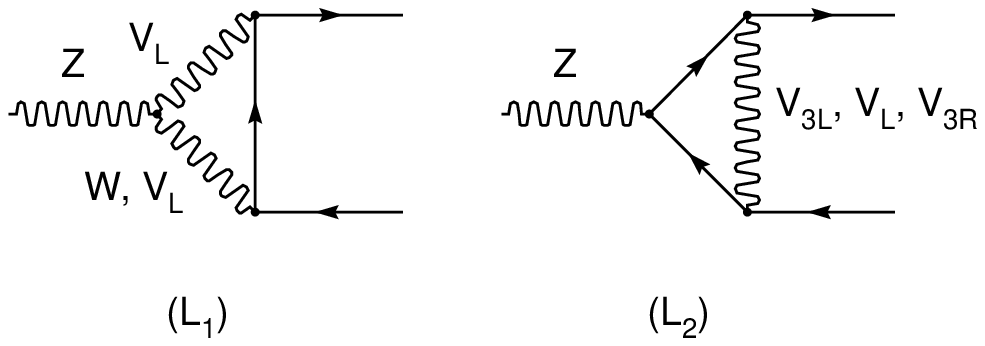}}
\noindent
\centerline{\bf Fig. 14}
\end{figure}

\begin{figure}
\epsfysize=6truecm
\centerline{\epsffile{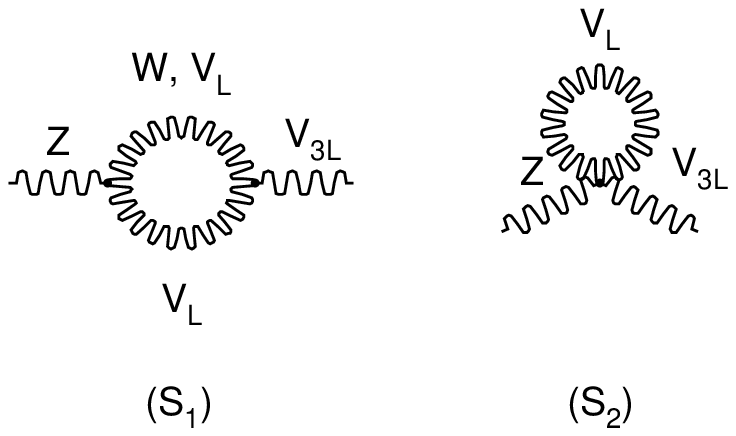}}
\noindent
\centerline{\bf Fig. 15}
\end{figure}

\begin{figure}
\epsfysize=6truecm
\centerline{\epsffile{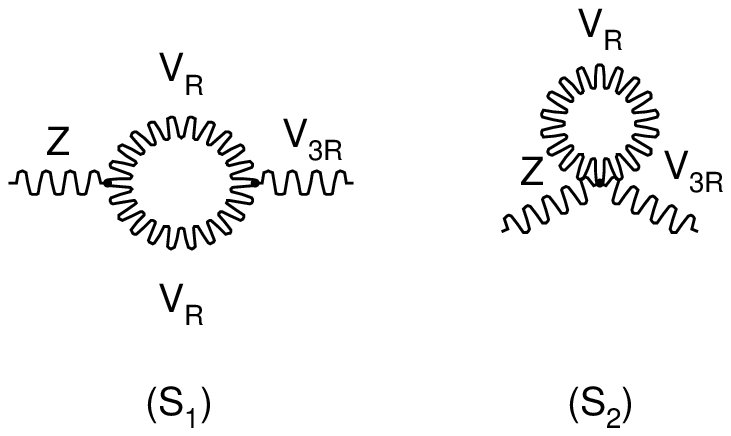}}
\noindent
\centerline{\bf Fig. 16}
\end{figure}
\newpage

\end{document}